\documentclass[smallextended]{svjour3}       
\smartqed  
\usepackage{graphicx,natbib,amsmath,xcolor,hyperref}
\usepackage{float}
\usepackage{caption}
\usepackage{threeparttable}
\usepackage{booktabs}
\newcommand{\diff}{\text{\rm d}}
\newcommand{\figref}[1]{Fig.~\ref{#1}}

\newcommand{\BC}{{\boldsymbol{\mathnormal C}}}
\newcommand{\BF}{{\boldsymbol{\mathnormal F}}}

\newcommand{\BM}{{\boldsymbol{\mathnormal M}}}

\newcommand{\superscr}[1]{\ensuremath{{}^{\rm #1}}}

\newcommand{\Bsigma} {\ensuremath{\boldsymbol\sigma}}

\newcommand{\Bomega  }{\ensuremath{\boldsymbol\omega}}

\newcommand{\Btau}   {\ensuremath{\boldsymbol\tau}}
\newcommand{\Bkappa} {\ensuremath{\boldsymbol\kappa}}

\newcommand{\Bb}{{\boldsymbol{\mathnormal b}}}

\newcommand{\Be}{{\boldsymbol{\mathnormal e}}}

\newcommand{\Bn}{{\boldsymbol{\mathnormal n}}}

\newcommand{\Br}{{\boldsymbol{\mathnormal r}}}
\newcommand{\Bs}{{\pmb{\mathnormal s}}}
\newcommand{\Bt}{{\boldsymbol{\mathnormal t}}}

\newcommand{\Bv}{{\boldsymbol{\mathnormal v}}}

\newcommand{\Balpha }{\ensuremath{\boldsymbol\alpha}}

\newcommand{\Beps    }{\ensuremath{\boldsymbol\epsilon}}

\newcommand{\Bepspl }{\ensuremath{\boldsymbol\epsilon\superscr{pl}}}
\newcommand{\Brho}{{\boldsymbol{\rho}}}
\newcommand{\Bmu}{{\boldsymbol{\mu}}}

\newcommand \MZ [1] {\bgroup\noindent[\textcolor{blue}{\textbf{MZ}: #1}]\egroup\ignorespacesafterend}
\newcommand \SG [1] {\bgroup\noindent[\textcolor{violet}{\textbf{SG}: #1}]\egroup\ignorespacesafterend}

\begin{document}

\titlerunning{Phase field model of dislocation microstucrure and recrystallization}
\title{A phase field model of the effects of dislocation microstructure on grain boundary motion during recrystallization}

\author{Yufan Zhang \and
        Michael Zaiser 
}


\institute{Y. Zhang and M. Zaiser\at
           Friedrich-Alexander Universit\"at Erlangen-N\"urnberg \\
           Department of Materials Science, WW8-Materials Simulation\\
           Dr.-Mack Strasse 77, 90762 F\"urth, Germany
           \email{shucheta.shegufta@fau.de}           
}

\date{Received: date / Accepted: date}

\maketitle

\begin{abstract}
The internal energy associated with the defect microstructure of strongly deformed crystals provides an important driving force for grain boundary motion during recrystallization. Typical dislocation microstructures are strongly heterogeneous and this heterogeneity affects the motion of recrystallization boundaries. In this study, a phase field model for microstructure evolution encompassing the evolution of both dislocation densities and grain order parameters is formulated. The model is employed to generate typical dislocation microstructures exhibiting multiscale features such as incidental and geometrically necessary dislocation walls. It is then used to study the motion of recrystallization boundaries in the associated complex defect energy 'landscape'. Results are compared to experimental observations.

\end{abstract}

\maketitle


\section{\label{intro} Introduction}

Phase field models have become an important tool for simulation of grain growth and grain microstructure evolution (for a general overview of phase field models for microstructure, see \citet{steinbach2009phase}). There exist a vast number of phase field models for grain microstructures which use different types of order parameters. Multi-phase- field models (MPF) use for each grain a separate order parameter (see e.g. \citet{fan1997computer,kim2006computer,moelans2008quantitative,moelans2009comparative}); 
while orientation-field models use a single phase field order parameter characterizing the local degree of crystalline order, and couple this to an orientation field (see e.g. \citet{kobayashi2000continuum,henry2012orientation}). It has been observed that, for large polycrystalline aggregates, MPF models may require a large number of order parameters, necessitating re-labeling schemes \cite{permann2016order}, and that they may need corrections to avoid emergence of spurious grain boundary phases \citep{toth2015consistent}. Both problems are not an issue in the present study, where we focus on the influence of dislocation microstructure on the motion of a single grain boundary. We therefore use the MPF approach in the formulation of \citet{moelans2008quantitative} as starting point for the present investigation. 

\begin{figure}[htpb]  \centering
    \includegraphics[width = .9\textwidth]{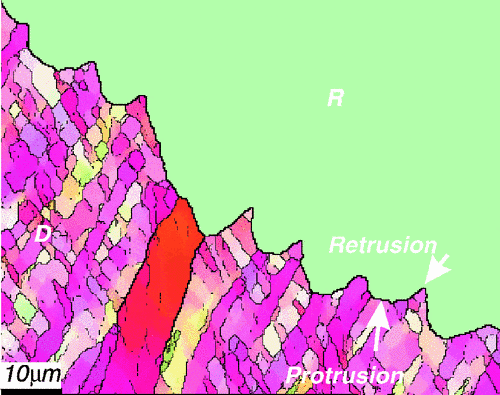}
    \caption{Micrograph showing the shape of a recrystallization boundary migrating into a heterogeneous microstructure of geometrically necessary and incidental dislocation boundaries; pure aluminum cold rolled to 50\% thickness reduction at room temperature, the letters R and D refer to recrystallized and deformed, respectively, the colors indicate local lattice orientation; after \citet{moelans2013phase}.}
    \label{fig:recrystallization}
\end{figure}
It is generally accepted that an important thermodynamic driving force for recrystallization is provided by the mechanical work stored in form of plasticity-induced defects ('defect energy'). These defects re-arrange or annihilate during recrystallization 
\citep{doherty1997recrystallization}, and the recovered energy may drive the motion of grain boundaries. After prolonged deformation, the majority defect energy resides in strongly heterogeneous dislocation microstructures consisting of 'geometrically necessary' (i.e., misoriented) and 'incidental' dislocation walls \cite{hughes2003geometrically}.    dislocations. The defect energy of these microstructures may alternatively be expressed in terms of the spacing and energy of dislocation boundaries, plus a contribution of statistically stored dislocations \citep{Doherty2001primary}, or equivalently in terms of the dislocation line energy of geometrically necessary and statistically stored dislocations (note that the energies of small-angle grain boundaries are equal to the energy of the dislocations contained therein).

The strong heterogeneity of the defect microstructure implies strong spatial fluctuations of the thermodynamic driving force for grain boundary motion, which superimpose on the effects of grain boundary energy and orientation dependent grain boundary mobility to induce complex, wavy patterns on recrystallization fronts as illustrated in Figure \ref{fig:recrystallization}. In a series of papers, Moelans and co-workers \citep{moelans2013phase,yadav2021effects,yadav2021influence} generalized the phase field model of \citet{moelans2008quantitative} by including a space dependent 'deformation energy' to mimic the spatially inhomogeneous distribution of defects, and studied how variations in the spatial distribution of defects including presence of structures with multiple wavelengths affect the motion and morphology of a recrystallization front. \citet{hamed2022impact} used a similar approach  where they model the heterogeneity of dislocation microstructures by assigning localized energy values to synthetic dislocation microstructure realizations that mimic experimental data, and they studied the expansion of recrystallized grains from localized nuclei within these synthetic structures. One drawback of these previous studies is that they use phenomenological frameworks to model the spatial modulations of the defect energy, which are simply introduced as prescribed energy functions that provide the initial conditions for a model of grain boundary motion that includes degradation of the defect energy. Although the parameters of these functions may be related phenomenologically to experimental scaling laws \citep{hamed2022impact}, it would be more desirable to obtain the spatial fluctuations of the defect energy (including measures such as dislocation density, dislocation wall spacing, and local orientation/misorientation) from a theoretical model. This is the purpose of the present study.

A useful starting point for such an investigation is provided by the phase field models of dislocation pattern evolution in one and two dimensions wthat ere formulated by Groma, Zaiser and co-workers \cite{groma2016dislocation,wu2018instability,wu2021cell}. An important feature of these models is a defect energy function which describes the energy of the dislocation systems in terms of dislocation density variables (see also \citet{zaiser2015local,berdichevsky2016energy}). Together with the  energy associated with the elastic strain field, this energy provides the thermodynamic driving force for dislocation microstructure evolution. Thus, the formulation and evolution of a defect energy functional are integral components of the constitutive framework. While these models were originally formulated in 2D, assuming straight parallel dislocations, \citet{groma2021dynamics} built on the continuum theory of dislocation kinematics by \citet{hochrainer2014continuum} to provide an analogous theoretical model for the evolution of dislocation densities on a single slip system in three dimensions. Finally, \citet{zhang2025continuum} generalized this model to multiple slip conditions and applied it to the evolution of dislocation microstructures in 3D crystals. They demonstrated that this model captures the spontaneous emergence of two-scale dislocation patterns associated with lattice misorientations, in the terminology of \citet{hughes2003geometrically} representing 'incidental' and 'geometrically necessary' dislocation walls. In the present investigation, we generalize this model to larger strains and apply it to the evolution of multi-scale dislocation and lattice orientation patterns. We then couple the model to a three-dimensional phase field model of grain microstructure evolution and use it to investigate the morphology of recrystallization fronts as function of the orientation of the front and of microstructure parameters. 

 Section 2 of this study presents the employed phase field model for the evolution of dislocation densities and the model for evolution of grain order parameters. In Section 3.1 we apply these models to simulate the evolution of dislocation microstructures in B2 (rocksalt) lattices including the concomitant three-dimensional lattice orientation patterns, and in Section 3.2 we study the motion of recrystallization fronts in the same structures. A discussion of the results is provided in Section 4, alongside an outlook on future perspectives for the present approach.

\section{Constitutive framework}

\subsection{Dislocation microstructure evolution}

The constitutive framework for dislocation microstructure evolution represents a dislocation based crystal plasticity model with dislocation transport. This requires the formulation of kinematic equations describing the kinematic evolution of the crystal manifold and the evolution of dislocation densities, and the derivation of thermodynamic forces driving this evolution

\subsubsection{Continuum kinematics and internal variables}

We use a small strain approximation where we additively decompose the deformation gradient into elastic and plastic parts, $\BF = \BF^{\rm el} + \BF^{\rm pl} = \BF^{\rm el} + \Bepspl + \Bomega^{\rm pl}$ where $\Bepspl$ and $\Bomega^{\rm pl}$ are plastic strain and rotation tensors.  

Plastic deformation is envisaged in terms of a set of slip system specific shear strains $\gamma{\beta}(\Br)$ where $\beta$ is a slip system  index. The slip systems are characterized by unit slip plane normal vectors $\Bn^{\beta}$ and slip vectors $\Bs^{\beta}$; the corresponding Burgers vectors are $\Bb^{\beta} = b \Bs^{\beta}$. The overall plastic strain tensor is obtained from the slip system variables as
\begin{equation}
    \Bepspl = \sum_{\beta} \gamma^{\beta} M^{\beta}
\end{equation}
where the slip system projection tensors are given by
\begin{equation}
    M^{\beta} = \frac{1}{2}(\Bn^{\beta} \otimes \Bs^{\beta} + \Bs^{\beta} \otimes \Bn^{\beta})
\end{equation}
where $\otimes$ denotes the tensor product. The plastic rotation tensor is given by
\begin{equation}
    \Bomega^{\rm pl} = \sum_{\beta} \gamma^{\beta} W^{\beta}
\end{equation}
with the slip system rotation tensors
\begin{equation}
    W^{\beta} = \frac{1}{2}(\Bn^{\beta} \otimes \Bs^{\beta} - \Bs^{\beta} \otimes \Bn^{\beta})
\end{equation}

Dislocation microstructure is described in terms of space dependent, slip system specific dislocation densities $\rho^{\beta}(\Br)$ which locally represent the length of dislocation lines per unit volume (irrespective of orientation) in the slip system $\beta$. Geometrically necessary dislocations are characterized by a set of dislocation density vectors $\Brho^{\beta}(\Br)$, which can be obtained from the local shear strains via
\begin{equation}
    \Brho^{\beta}(\Br) = - (\Bn^{\beta} \times \nabla) \gamma^{\beta}.
\end{equation}
These vectors can be used to construct the Kr\"oner-Nye dislocation density tensor as
\begin{equation}
    \Balpha(\Br) = - \nabla \times \Bepspl =
    \sum_{\beta} \Brho^{\beta} \otimes \Bb^{\beta}.
\end{equation}

\subsubsection{Kinematics of dislocation densities}

Our kinematic framework describes a system where dislocations of different orientation may have different velocities. Specifically, we assume the velocity function to depend on the tangent vector via
\begin{equation}
    v(\Bt) = v_0 + v_{1} (\Bt\cdot \Bt^{\beta})
    \label{eq:v12}
\end{equation}
where $\Bt^{\beta} = \Brho^{\beta}/|\Brho^{\beta}|$. This form is equivalent to the one assumed by Groma and co-workers \cite{groma2021dynamics}.
From the dislocation density variables, the plastic slip rates are evaluated using the generalized Orowan relation \cite{groma2021dynamics,zhang2025continuum}
\begin{equation}
\dot{\gamma}^{\beta} 
= b (\rho^{\beta} v_0^{\beta} + \Brho^{\beta}\cdot \Bv^{\beta}_1),
\end{equation}
where we have introduced the notation $\Bv^{\beta}_1 = \Bt^{\beta} v^{\beta}_{1}$. Accordingly, the evolution equation of the GND density vector is
\begin{equation}
\partial_t \Brho^{\beta} = - [\Bn^{\beta} \times \nabla] [\rho^{\beta} v_0^{\beta} + \Brho^{\beta}\cdot \Bv^{\beta}_1]
\end{equation}
For the scalar dislocation densities the evolution equation (for derivation see \citet{zhang2025continuum}) is given by
\begin{eqnarray}
\partial_t \rho^{\beta} &=& - [\Bn^{\beta} \times \nabla] \cdot [\Brho^{\beta} v^{\beta}_0 + \rho^{\beta}\Bv^{\beta}_1]\nonumber\\
&+& q^{\beta}_0 v^{\beta}_0 + \frac{1-\Phi^{\beta}}{2}\rho^{\beta}[\Bn^{\beta} \times \nabla] \cdot\Bv^{\beta}_1 +
\Phi^{\beta}\rho^{\beta} k^{\beta}_{\kappa} \Bt^{\beta}_{\kappa} \cdot \Bv^{\beta}_1.
\label{eq:rhomulti}
\end{eqnarray}
In this equation, the anisotropy function $\Phi^{\beta}$ is evaluated as
(for derivation and discussion see \citet{monavari2014comparison,monavari2016continuum})
\begin{equation}
\Phi^{\beta}(\rho^{\beta}, \Brho^{\beta}) \approx \frac{1}{2}\left[\left( \frac{\Brho^{\beta}\cdot \Brho^{\beta}}{(\rho^{\beta})^2}\right)+\left( \frac{\Brho^{\beta}\cdot \Brho^{\beta}}{(\rho^{\beta})^2}\right)^3\right].
\end{equation}
The GND curvature is evaluated as
\begin{equation}
k_{\Brho}^{\beta} = [\Bn^{\beta} \times \nabla]\cdot\Bt^{\beta}.
\end{equation}
and the average curvature as
\begin{equation}
    q^{\beta}_0 = \rho^{\beta} [\Phi k_{\Brho}^{\beta}  + (1-\Phi) k^{\beta}_{\rm SSD}]
    \quad,\quad k^{\beta}_{\rm SSD} = \xi \;\rm{sign}(\tau^{\beta})\sqrt{\rho}
    \label{eq:kbeta}
\end{equation}
This implies that the curvature radii of SSD loops are assumed proportional to the mean dislocation spacing, and the loops are oriented such that they expand under the shear stress acting on the respective slip system. 

\subsubsection{Free energy functional and thermodynamic driving forces}

We write the free energy of the deforming crystal as 
\begin{equation}
F = \int_V (\Psi_{\rm E} + \Psi_{\rm D}) \diff V
\end{equation}
where the elastic energy density is given by 
\begin{equation}
\Psi_{\rm E} = \frac{1}{2} (\Beps - \Bepspl) : \BC : (\Beps - \Bepspl)
\end{equation}
and $\BC$ is the tensor of elastic constants which we take to represent an isotropic material of shear modulus $G$ and Poisson ratio $\nu$. The evolution of the elastic energy density is controlled by the plastic strain rate $\dot{Bepspl} = \sum_{\beta} \BM^{\beta} \dot{\gamma}\beta$. 

The defect energy is assumed in the form \citep{zaiser2015local,zhang2025continuum}
\begin{equation}
\Psi_{\rm D} = \sum_{\beta}\left(A G b^2 \rho^{\beta} \ln (\rho {\ell}^2) + D G b^2 \frac{\Brho^{\beta}\cdot \Brho^{\beta}}{2 \rho}\right).
\label{eq:energycdd2}
\end{equation}
where $A$ and $D$ are numerical constants of the order of unity and $\rho = \sum_{\beta}\rho^{\beta}$. Associated with the free energy density, we define quasi-chemical potentials for the dislocation density variables: 
\begin{eqnarray}
\mu^{\beta}_{\rho} = \frac{\partial \Psi_{\rm D}}{\partial \rho^{\beta}} &= G b^2 \left[A[\ln(\rho {\ell}^2)+1] - B \frac{\Brho^{\beta}\cdot \Brho^{\beta}}{2\rho^2}\right],\nonumber\\
\Bmu^{\beta}_{\Brho} = \frac{\partial \Psi_{\rm D}}{\partial \Brho^{\beta}} &= D G b^2  \frac{\Brho^{\beta}}{\rho},
\end{eqnarray}  
where $\Bmu^{\beta}_{\Brho}$ is a vector compiling the potentials $(\mu^{\beta}_{\Brho})_i = \partial \Phi_{\rm D}/\partial \Brho^{\beta}_i$. Moreover, we define the resolved shear stresses on the slip systems as 
\begin{equation}
    \tau^{\beta} = \BM^{\beta}:\BC:(\Beps - \Bepspl)
\end{equation}
and a set of defect microstructure related shear stress-like quantities as 
\begin{eqnarray}
\tau^{\beta}_{\Brho} &=& - (\Bn^{\beta} \times \nabla) \Bmu^{\beta}_{\Brho}
\;,\; 
\Btau^{\beta}_{\rho,0} = - (\Bn^{\beta} \times \nabla) \mu^{\beta}_{\rho},\nonumber\\
\Btau^{\beta}_{\rho,1} &=& \Btau^{\beta}_{\rho,0} + \displaystyle \frac{1}{2\rho^{\beta}} (\Bn^{\beta} \times \nabla)[ (1-\Phi^{\beta})\rho^{\beta}\mu^{\beta}_{\rho}], \nonumber\\
\tau^{\beta}_{\rm c} &=& \displaystyle \frac{\mu_{\rho}^{\beta} q^{\beta}_0}{\rho^{\beta}}\;,\;
\Btau^{\beta}_{\rm c} = \Phi^{\beta}\rho^{\beta} k^{\beta}_{\Brho} \Bt^{\beta}_{\Brho}.
\end{eqnarray} 
The terminology here is such that stresses associated with gradients of the chemical potential $\mu_{\Brho}$ of the GND density vector carry the subscript $\Brho$, and stresses associated with the chemical potential $\mu_{\rho}$ of the scalar total dislocation density carry the subscript $\rho$. 
With these notations, the evolution of the free energy function can be written as
\begin{eqnarray}
\frac{{\diff} F}{{\diff} t} =  \sum_{\beta} \int_V & \left\{- (\tau^{\beta} +\tau^{\beta}_{\Brho})  
[\rho^{\beta} v^{\beta}_0 + \Brho^{\beta}\cdot \Bv^{\beta}_1] - \Btau_{\rho,0}^{\beta} \cdot
[\Brho^{\beta} v^{\beta}_0 + \rho^{\beta} \Bv^{\beta}_1] \right. \nonumber\\
& \left. + \rho^{\beta} \tau^{\beta}_{\rm c} v^{\beta}_0 b + \rho^{\beta} (\Btau_{\rho,1}-\Btau_{\rho,0}) 
\cdot \Bv^{\beta}_1 b + \rho^{\beta} \Btau^{\beta}_{\rm c} \cdot \Bv^{\beta}_1 b \right\}\diff V.
\end{eqnarray}
Thermodynamic consistency requires the term in curved brackets to be negatively definite. Sorting the terms one finds that this condition reads 
\begin{eqnarray}
&\left(\tau^{\beta} +\tau^{\beta}_{\Brho} + \frac{\Brho^{\beta}}{\rho^{\beta}}\cdot \Btau_{\rho,0}^{\beta} -
\tau^{\beta}_{\rm c} \right)v^{\beta}_0 \nonumber\\
+ &\left(\frac{\Brho^{\beta}}{\rho^{\beta}}(\tau^{\beta} +\tau^{\beta}_{\Brho}) -\Btau^{\beta}_{\rho,1} - \Btau_{\rm c}^{\beta} \right) \cdot \Bv^{\beta}_1 \ge 0
\label{eq:TDconmult1}
\end{eqnarray}

While any velocity law that is thermodynamically consistent must match this condition, there is no unique rule for defining such consistent laws. Here we adopt the velocity functions analyzed by \citet{groma2021dynamics}, who match the velocity laws derived for 2D dislocation systems using direct averaging of the Peach-Koehler forces \citep{groma2016dislocation}. These velocity functions are givren by
\begin{equation}
v_0^{\beta} = \frac{B}{b}\chi(\tau^{\beta}_*,\tau^{\beta}_{\rm y})\quad,\quad
v_1^{\beta} = \frac{B}{b}\Bt^{\beta}_{\kappa}\cdot \left(\frac{\Brho^{\beta}}{\rho^{\beta}}[1-\chi(\tau^{\beta}_*,\tau^{\beta}_{\rm y})] + \Btau^{\beta}_{\rho,1}\right),
\label{eq:TDyield3}
\end{equation}
where $\tau^* = \tau^{\beta} + \tau^{\beta}_{\Brho}$ and $\tau^{\beta}_{\rm y}$ is the positively definite slip system yield stress ('friction stress'). The function $\chi(\tau^{\beta}_*,\tau^{\beta}_{\rm y})$ is defined via 
\begin{equation}
\chi(\tau^{\beta}_*,\tau^{\beta}_{\rm y}) = \left\{
\begin{array}{ll}
\tau^{\beta}_*- \tau^{\beta}_{\rm y},& \tau^{\beta}_*>  \tau^{\beta}_{\rm y}; \\
- \tau^{\beta}_*+ \tau^{\beta}_{\rm y},& \tau^{\beta}_*< - \tau^{\beta}_{\rm y};\\
0 & {\rm otherwise}
\end{array}\right.
\end{equation}
We note that the slip system yield stress $\tau^{\beta}_{\rm y}$ must be chosen such that in the flowing phase the driving force $\tau^*$ is always high enough to ensure that the plastic work exceeds the energy required for the expansion of the dislocation loop \citep{wu2022thermodynamic}; this leads to a constraint of the form $\tau^{\beta}_{\rm y} \ge \eta (\mu_{\rho}^{\beta}/b) \sqrt{\rho}$. 

\subsection{Constitutive framework for recrystallization}

The constitutive framework for the recrystallization process describes the effects of the heterogeneous dislocation microstructure on the growth of a recrystallized grain. This requires a phase-field model, in which the free energy functional couples order parameters characterizing the grain and dislocation microstructure, together with specification of kinematic equations that govern the evolution of the grain order parameters. As starting point we use the phase-field model for recrystallization introduced by Moelans et al.\citep{moelans2013phase}.This model considers two order parameters, $\phi_{\rm rex}$ representing the recrystallized, defect free lattice and $\phi_{\rm def}$ representing the initial crystal which, after deformation, contains an inhomogeneous defect microstructure. The state $\phi_{\rm rex} = 1$ and $\phi_{\rm def} = 0$ corresponds to the recrystallized region, whereas $\phi_{\rm rex} = 0$ and $\phi_{\rm def} = 1$ corresponds to the lattice of the initial, deformed crystal.
The free energy functional of the model is given by,
\begin{equation}
	\label{eq:Fgrain}
	F=\int_V (\Psi_{\rm G} + \Psi_{\rm D}) dV,
\end{equation}
where $\Psi_{\rm G}$ and $\Psi_{\rm D}$ represent the respective contributions of the grain boundary energy and the stored energy associated with defects.
The formulation of $\Psi_{\rm G}$ is based on Chen and Yang’s grain growth model\citep{chen1994computer},
\begin{eqnarray}
	\Psi_{\rm G} &=& \frac{6\sigma_{\rm g}}{l_{\rm g}} \left(\frac{\phi_{\rm rex}^4}{4}+\frac{\phi_{\rm def}^4}{4} - \frac{\phi_{\rm rex}^2}{2}-\frac{\phi_{\rm def}^2}{2} + \frac{3}{2}\phi_{\rm rex}^2\phi_{\rm def}^2+\frac{1}{4}\right)\nonumber\\ 
	&+& \frac{3}{8}\sigma_{\rm g}l_{\rm g}\left((\nabla \phi_{\rm rex})^2) +  (\nabla \phi_{\rm def})^2\right),
\end{eqnarray}
where $\sigma_{\rm g}$ is the grain boundary energy and $l_{\rm g}$ is the diffuse grain boundary width.
The thermodynamic driving force for grain boundary migration is provided by grain boundary energy minimization on the one hand, and by reduction of the stored defect energy on the other hand. The idea is that the re-organization of the crystal lattice that is associated with grain boundary migration leads to an elimination of defects. To describe this process, we introduce the function
\begin{equation}
f = \frac{\phi_{\rm def}^2}{\phi_{\rm rex}^2+\phi_{\rm def}^2}
\end{equation}
which may be interpreted as the local fraction of deformed crystal lattice \citep{moelans2011quantitative}. If this fraction decreases, then this leads to a degradation  of the defect densities $\{\rho^{\beta},\Brho^{\beta}\}$, which we describe by the equations
\begin{eqnarray}
    \frac{\partial_t \rho^{\beta}}{\rho^{\beta}} &=& 
    \left\{
    \begin{array}{l} \displaystyle\frac{\partial_t f}{\check{f}} \quad,\quad f = \check{f} \quad {\rm and} \quad \dot{f} < 0\nonumber\\
    0 \quad{\rm otherwise} 
    \end{array}\right. ,\quad {\rm where}\quad  
    \check{f}=\min_{t' \le t} f(t)
    \\
\frac{\partial_t (\Brho^{\beta}.\Brho^{\beta})}{\Brho^{\beta}.\Brho^{\beta}}
 &=& \frac{\partial_t \rho^{\beta}}{\rho^{\beta}}\left(2   - \frac{2 A}{nD}  \frac{\rho^2}{\Brho^{\beta}.\Brho^{\beta}} \right). 
 \label{eq:disdegrad}
\end{eqnarray}
Here, $n$ is the number of active slip systems. We show in Appendix A that, when the change in dislocation densities is exclusively due to the change in $f$, these equations can be integrated to yield 
\begin{equation}
	\label{eq:fd}
	\Psi_{\rm D}(t) = \check{f}(t) \Psi_{\rm D}^0.
\end{equation}
where $\Psi_{\rm D}^0$ is the defect energy density before the onset of recrystallization. This is the form used by \citet{moelans2013phase}. Our formulation emphasizes that it is not merely the defect energy that changes due to the change in crystallographic order parameters, but also the entire defect microstructure. 

The evolution equations for the grain order parameters simply represent Allen-Cahn type equations derived from the free-energy functional $F$,
\begin{equation}
	\label{eq:acrex}
	\partial_t \phi_{\rm rex} = -L\frac{\delta F}{\delta \phi_{\rm rex}},
\end{equation}
and
\begin{equation}
	\label{eq:acdef}
	\partial_t \phi_{\rm def} = -L\frac{\delta F}{\delta \phi_{\rm def}}.
\end{equation}
Following \citet{moelans2013phase}, the kinetic coefficient is taken as $L = 4M_{\rm g}/3l_{\rm g}$, where $M_{\rm g}$ is the grain boundary mobility. By substituting Eq.(\ref{eq:Fgrain})-Eq.(\ref{eq:fd}) into Eq.(\ref{eq:acrex}) and Eq.(\ref{eq:acdef}), we obtain
\begin{eqnarray}
	\partial_t \phi_{\rm rex} &=& -L\left[\frac{6\sigma_{\rm g}}{l_{\rm g}} (\phi_{\rm rex}^3 - \phi_{\rm rex}^2 + 3\phi_{\rm rex}\phi_{\rm def}^2) - \frac{3}{4}\sigma_{\rm g}l_{\rm g}\nabla^2 \phi_{\rm rex}-\frac{2\phi_{\rm rex}\phi_{\rm def}^2
    \Psi_{\rm D}^0}{(\phi_{\rm rex}^2+\phi_{\rm def}^2)^2}\right]\nonumber\\
	\label{eq:dpdt}
	\partial_t \phi_{\rm def} &=& -L\left[\frac{6\sigma_{\rm g}}{l_{\rm g}} (\phi_{\rm def}^3 - \phi_{\rm def}^2 + 3\phi_{\rm def}\phi_{\rm rex}^2) - \frac{3}{4}\sigma_{\rm g}l_{\rm g}\nabla^2 \phi_{\rm def}+\frac{2\phi_{\rm def}\phi_{\rm rex}^2
    \Psi_{\rm D}^0}{(\phi_{\rm rex}^2+\phi_{\rm def}^2)^2}\right].\nonumber\\
\end{eqnarray}

\subsection{Model Parameters}

We apply our constitutive framework to the evolution of dislocation patterns and to the motion of recrystallization fronts in crystals with B1 (rocksalt) lattice structure. While this choice is at first glance unusual, it reflects a personal preference and illustrates that dislocations do not exist only in fcc metals. We emphasize the strong analogies with dislocation behavior in fcc crystals; in fact it is straightforward to adapt the formalism to fcc lattices. Model parameters used in the following represent material properties of LiF and are given in \autoref{tab:parameters}.
\begin{table}[h]
    \centering
    \captionsetup{skip=4pt}
    \begin{threeparttable}
        \caption{Model parameters}
        \label{tab:parameters}
        {\renewcommand{\arraystretch}{1.2}
        \begin{tabular}{llll}
            \toprule
            \textbf{Parameter} & \textbf{Value} & \textbf{Unit} & \textbf{Reference} \\
            \midrule
            $C_{11}$ & 120 & GPa & \cite{jakata2018optical}\\
            $C_{12}$ & 58  & GPa & \cite{jakata2018optical}\\
            $C_{44}$ & 58  & GPa & \cite{jakata2018optical}\\
            $G$      & 58  & GPa & \cite{jakata2018optical}\\
            $A$      & 0.2 &  & \cite{zhang2025continuum} \\
            $D$      & 0.2 &  & \cite{zhang2025continuum} \\
            $\alpha$ & 0.3 &  & \cite{wu2018instability}\\
            $\ell$ & $1.3 \times 10^{-6}$ & m & \\
            $B$ & $3.38\times 10^{-6}$ & m$^{2}$GPa$^{-1}$s$^{-1}$ &\cite{fanti1969viscous} \\
            $b$ & 2.85 & \AA & \cite{rapoport2002imaging} \\
            $\rho_0$ & $10^{14}$ & m$^{-2}$ & \\
            $\sigma_{\rm g}$ & 0.4 & J\,m$^{-2}$ & \cite{spitzer1962intercrystalline} \\
            $l_{\rm g}$ & $10^{-7}$ & m & \cite{yadav2021effects}\\
            $L$ & $10^{-6}$ & m$^{3}$J$^{-1}$s$^{-1}$ & \cite{yadav2021effects}\\
            \bottomrule
        \end{tabular}}
    \end{threeparttable}
\end{table}

\section{Simulation results}

\subsection{Dislocation patterns, dislocation walls and lattice rotations}

We consider an infinite crystal with B1 lattice structure that is subject to a uni-axial stress $\Bsigma^{\rm ext} = \sigma \Be_x \otimes \Be_x$. We identify $\Be_x$ with the unit vector in the [100] lattice direction, i.e., we consider deformation in a high-symmetry orientation where there are four potentially active and two inactive slip systems. The slip vectors $\Bs^i = \Bb^i/b$ and slip plane normal vectors $\Bn^i$ of the potentially active slip systems are given by 
\begin{eqnarray}
        \Bs^1 = \frac{1}{\sqrt{2}}[110]\quad,\quad \Bn^1 = \frac{1}{\sqrt{2}}[1-10]\nonumber\\
        \Bs^2 = \frac{1}{\sqrt{2}}[1-10]\quad,\quad \Bn^2= \frac{1}{\sqrt{2}}[110]\nonumber\\    
        \Bs^3 = \frac{1}{\sqrt{2}}[101]\quad,\quad \Bn^3 = \frac{1}{\sqrt{2}}[10-1]\nonumber\\
        \Bs^4 = \frac{1}{\sqrt{2}}[10-1]\quad,\quad \Bn^4 = \frac{1}{\sqrt{2}}[101]\nonumber\\
        \label{eq:slipsystems}
\end{eqnarray}
The resolved shear stress on the active slip systems is $\tau^{\rm ext} = \sigma/M$ where $M = 2$. In the following we assume, until otherwise stated, a homogeneous initial dislocation density of $\rho(0) = 10^{14}$ m$^{-2}$ on each of the active slip systems. The initial deformation state is fully homogeneous, i.e., the GND vector densities are identically zero, $\Brho^i = 0 \forall i$. Model parameters are compiled in \autoref{tab:parameters}.
\begin{figure}[btp]  \centering
    \includegraphics[width = 0.9\textwidth]{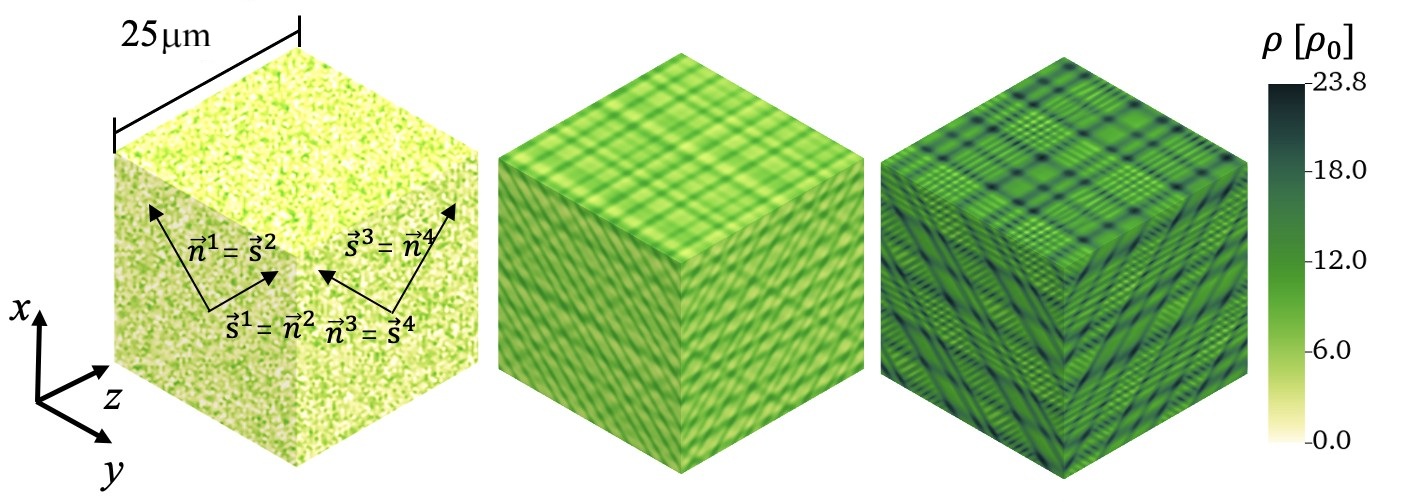}
    \caption{Formation of cellular patterns in symmetrical multiple slip of a LiF single crystal; total dislocation density at different shear strains.}
    \label{fig:cells4s}
\end{figure}
To study the formation of three-dimensional patterns, we consider symmetrical multiple slip conditions where we assume initially homogeneous slip activity on all 4 slip systems given in Eq. (\ref{eq:slipsystems}), i.e., all dislocation density vectors $\Brho^i = 0$, and initial dislocation densities of the form $\rho^i = \rho_0 + \delta \rho_i(\Br)$. For the perturbations required to initiate the patterning process, we now consider generic white-noise processes. Thus, we take the $\delta \rho^i$ to be Gaussian random fields with the correlation properties
\begin{equation}
\langle \delta \rho^i(\Br) \delta \rho^j(\Br')\rangle =
\rho_0^{2} \chi^2 \delta_{ij} \xi^3 \delta(\Br-\Br'),
\end{equation}
where the correlation length $\xi = \rho_0^{-1/2}$ is taken equal to the dislocation spacing. Technically, the random fields are implemented by assigning to each grid element $I$ with volume $\rho_0^{-3/2}$ a Gaussian distributed perturbation $\delta \rho(\Br_I) = \chi\rho_0 G_I$ where the $G_I$ are independent, standard normal distributed random variables.  In our simulations, we set $\chi = 0.01$. 
\begin{figure}[b]  \centering
    \includegraphics[width = \textwidth]{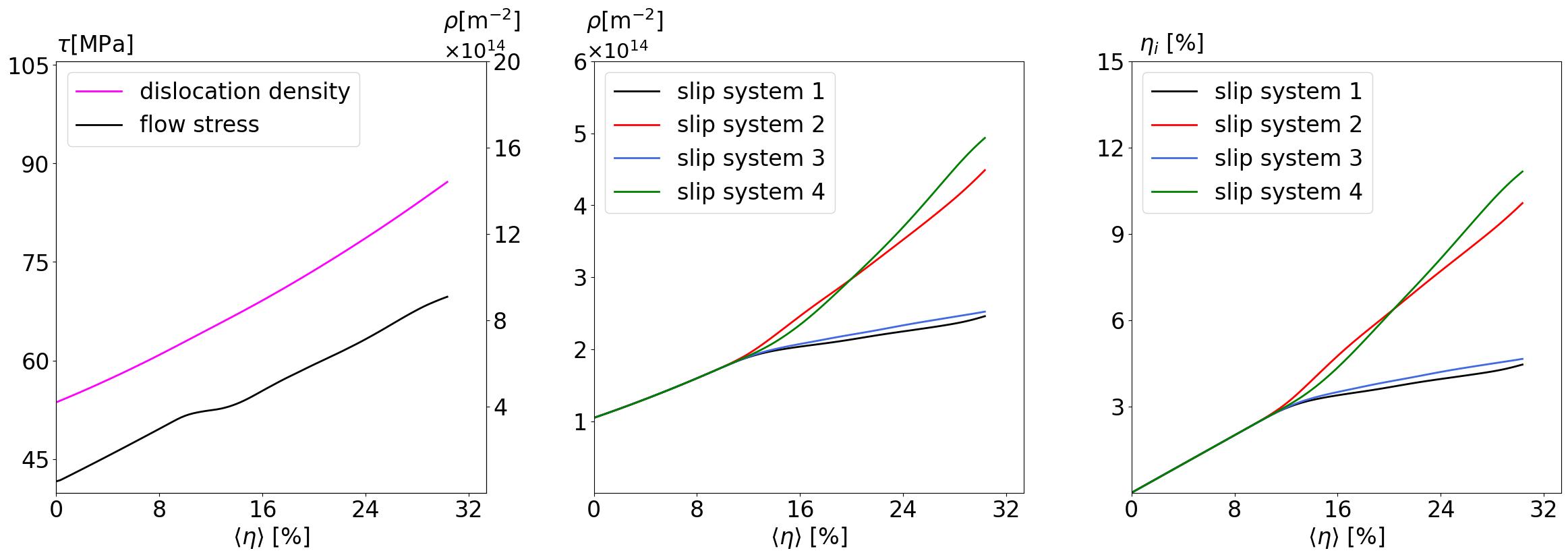}
    \caption{Evolution of stress, slip system strains, and dislocation densities during deformation in symmetrical multiple slip; left: total dislocation density and resolved shear stress on the 4 active slip systems; center: shear strains on the 4 active slip systems; right: dislocation densities on the active slip systems; all variables are given as functions of the average shear strain $\langle \gamma \rangle = M \epsilon_{xx} $ where $M=2$.}
    \label{fig:multislip}
\end{figure}

Dislocation microstructure evolution in 3D multi-slip conditions is characterized by the emergence of multiscale patterns. We use a large supercell with an edge length of $256 \rho_0^{-1/2}$ and periodic boundary conditions in all three dimensions. The size of the supercell corresponds to a physical scale of 25 $\mu$m. The patterning process is illustrated in Figure \ref{fig:cells4s}. During an initial stage up to a slip system strain of about 50$b\sqrt{\rho_0}$ (in dimensional units: 0.14, corresponding to an axial strain of 28\%), deformation is characterized by approximately symmetric activity on the 4 active slip systems. Wall patterns form on both pairs of conjugate slip systems and lead to a globally cellular structure (central graphs in Figure \ref{fig:cells4s}).  These walls carry near-zero net Burgers vector, i.e. in the terminology of \citet{hughes2003geometrically} they represent incidental dislocation boundaries. 

\begin{figure}[tb]  \centering
    \hspace*{.7cm}\hfill\includegraphics[width = .95\textwidth]{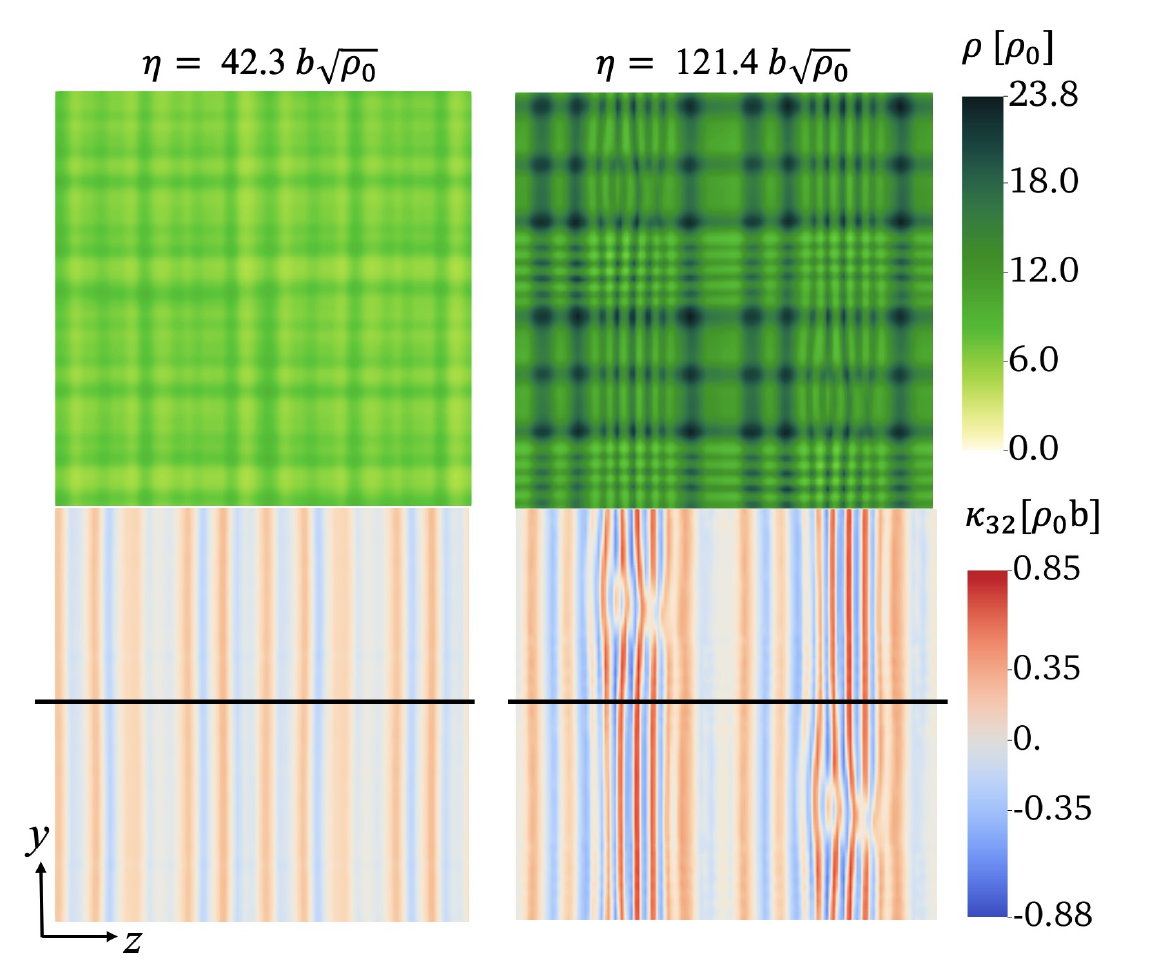}
    \includegraphics[width = .95\textwidth]{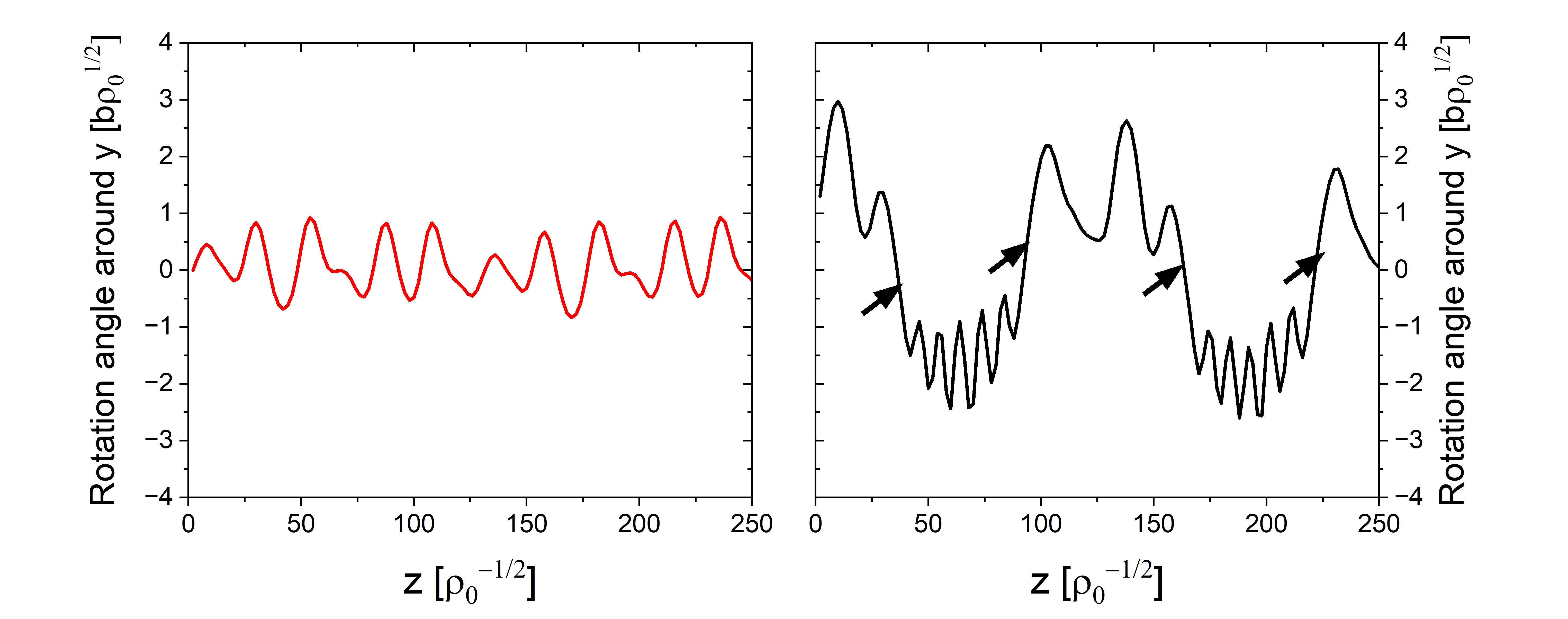}
    \caption{Dislocation arrangement in the [100] plane; 
    top: total dislocation density, center: lattice curvature tensor component $\kappa_{32}$, bottom: lattice orientation profiles along black line; 
    left: strain $\eta = 0.42 b\sqrt{\rho_0}$ (tensile strain 2.1\%) , right: slip system strain $\eta = 1.21 b\sqrt{\rho_0}$ (tensile strain 6.07\% ); the arrows in the right graph mark geometrically necessary dislocation walls. }
    \label{fig:boundaries}
\end{figure}
In a second stage, we observe a symmetry-breaking bifurcation with the consequence that slip activity on the conjugate slip systems becomes locally unbalanced. Symmetry breaking in the local slip activity pattern is accompanied by transformation of the dislocation arrangement into a two-scale pattern (right column in Figure \ref{fig:cells4s}). 
\begin{figure}[tb]  \centering
    \includegraphics[width = .6\textwidth]{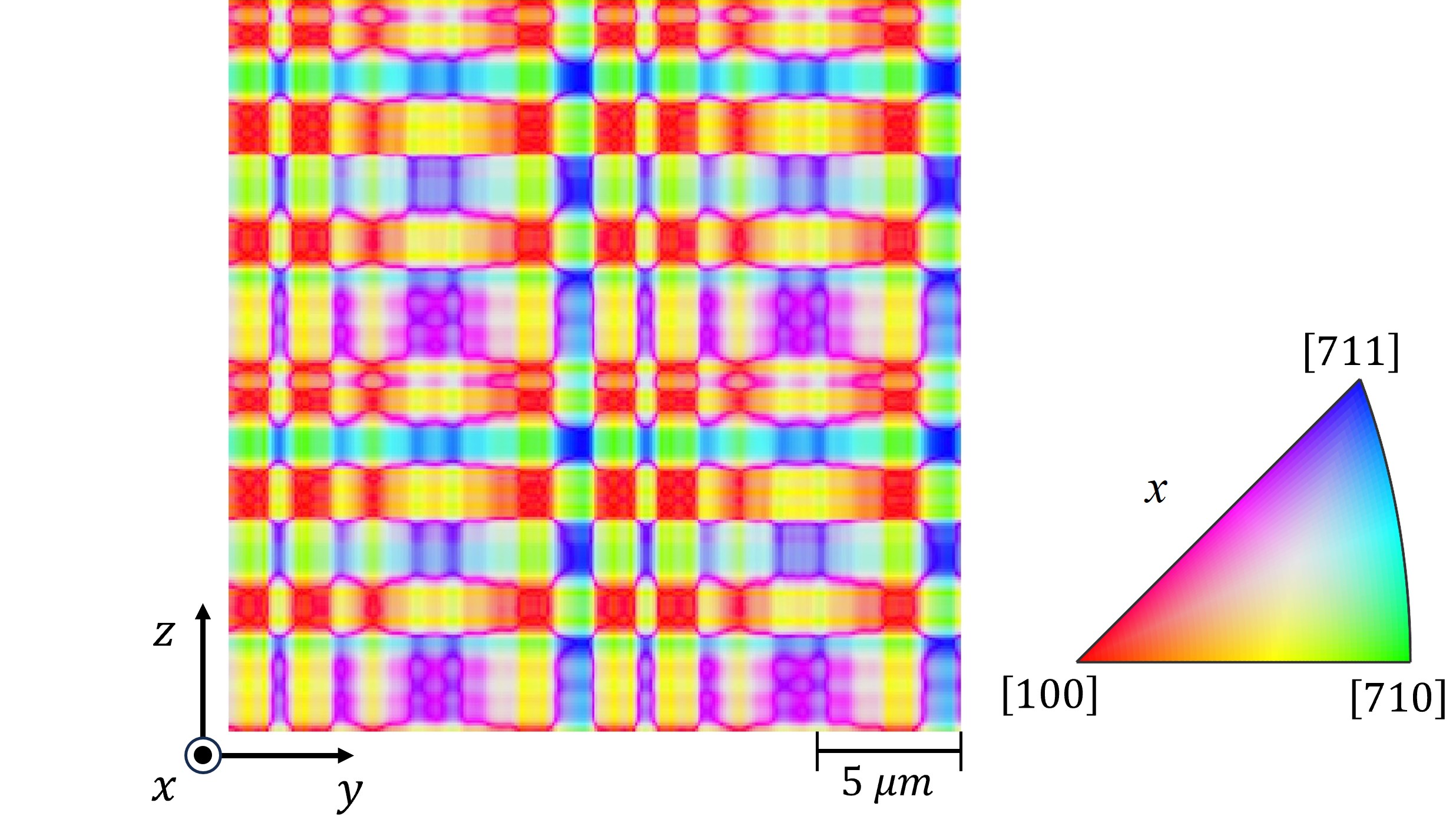}
    \includegraphics[width = .37\textwidth]{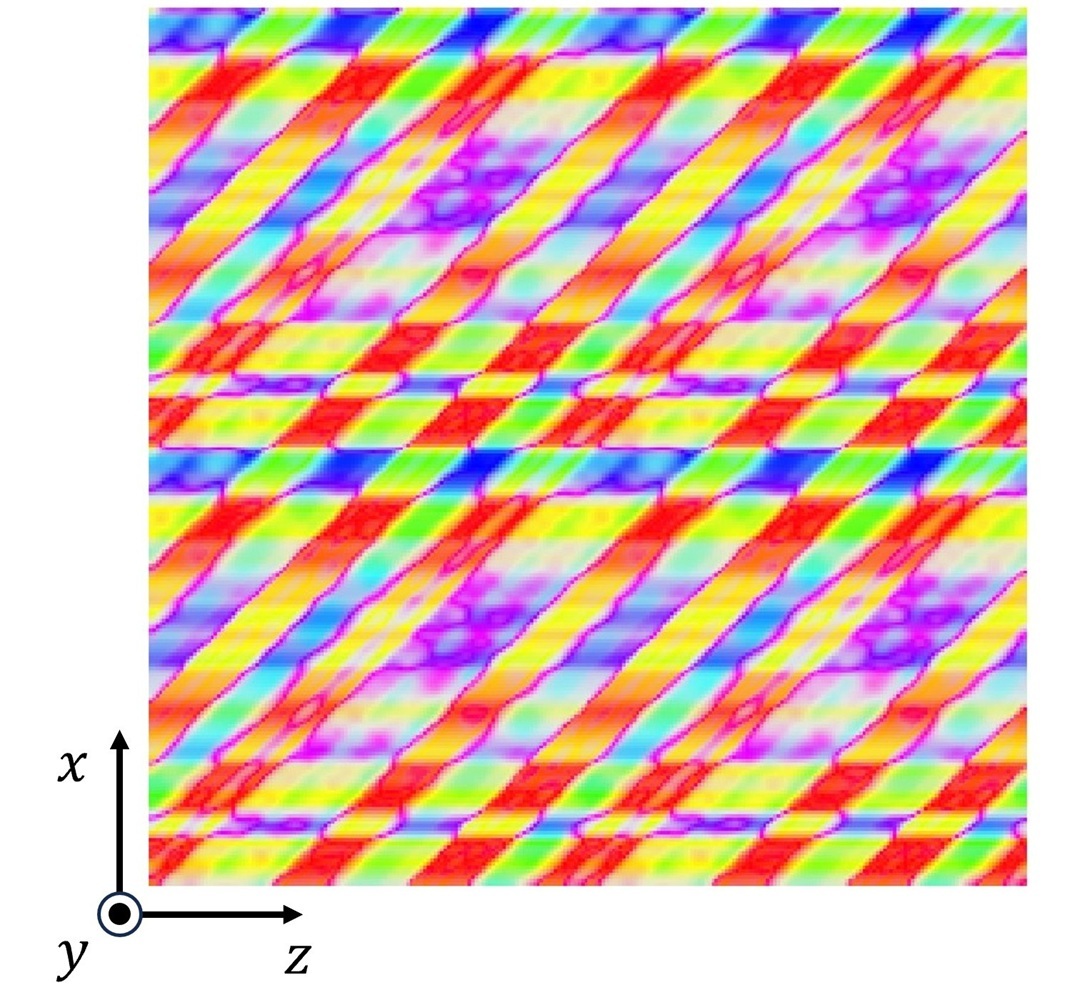}
    \caption{Lattice rotation patterns in planes perpendicular and parallel to the tensile axis, left: perpendicular plane, right: parallel plane; the colorscale indicates rotations towards the [110] pole as yellow-green and rotations towards the [111] pole as purple-blue. }
    \label{fig:orientations}
\end{figure}
To analyze this pattern in more detail, we take a closer look at the arrangement of dislocations in the plane perpendicular to the tensile axis. We use Nye's \cite{nye1953some} well known relation between lattice curvature tensor $\Bkappa$ and dislocation density tensor $\Balpha$, which for small deformations reads
\begin{equation}
\kappa_{ij} = \alpha_{ji} - \frac{1}{2}{\rm Tr}\Balpha \delta_{ij}.
\end{equation}
Before the bifurcation, we observe an arrangement of approximately equi-spaced dislocation walls, which are not associated with large-scale misorientations or spatial variations in slip system activity and accordingly carry no net GND content. In the terminology of \citet{hughes2003geometrically}, these dislocation walls classify as 'incidental' (Figure \ref{fig:boundaries}, left). After the bifurcation, two different types of dislocation walls are observed. Alongside with 'incidental' walls where lattice orientation and slip system activity on both sides are similar, there are walls which carry a significant lattice misorientation. Thus, in the terminology of Hughes and Hansen, these walls (marked by arrows in Figure \ref{fig:boundaries}, left) classify as 'geometrically necessary' since they separate regions of different lattice orientation and slip system activity. The ensuing lattice orientation patterns are illustrated in \figref{fig:orientations}: While some areas of the crystals remain near the initial [100] orientation, others develop alternating misorientations which tilt them towards the [110] and [111] poles. At the strains accessible by the present model, which uses a small-strain formulation, the tilt angles are still comparatively small. Nevertheless we shall see that the complex pattern of geometrically necessary and statistically stored dislocations has a strong influence on recrystallization. 

\subsection{Motion of recrystallization fronts}

We now consider the motion of grain boundaries in a static recrystallization scenario, i.e., we remove the applied loading such that the dislocation microstructure remains at rest. We then introduce a nucleus of a recrystallized grain. First, in order to study motion of planar recrystallization fronts, we introduce the nucleus in form of a thin system spanning planar lamella along which we set $\phi_{\rm rex} = 1, \phi_{\rm def} = 0$. 
\begin{figure}[tb]  
    \centering\includegraphics[width = .39\textwidth]{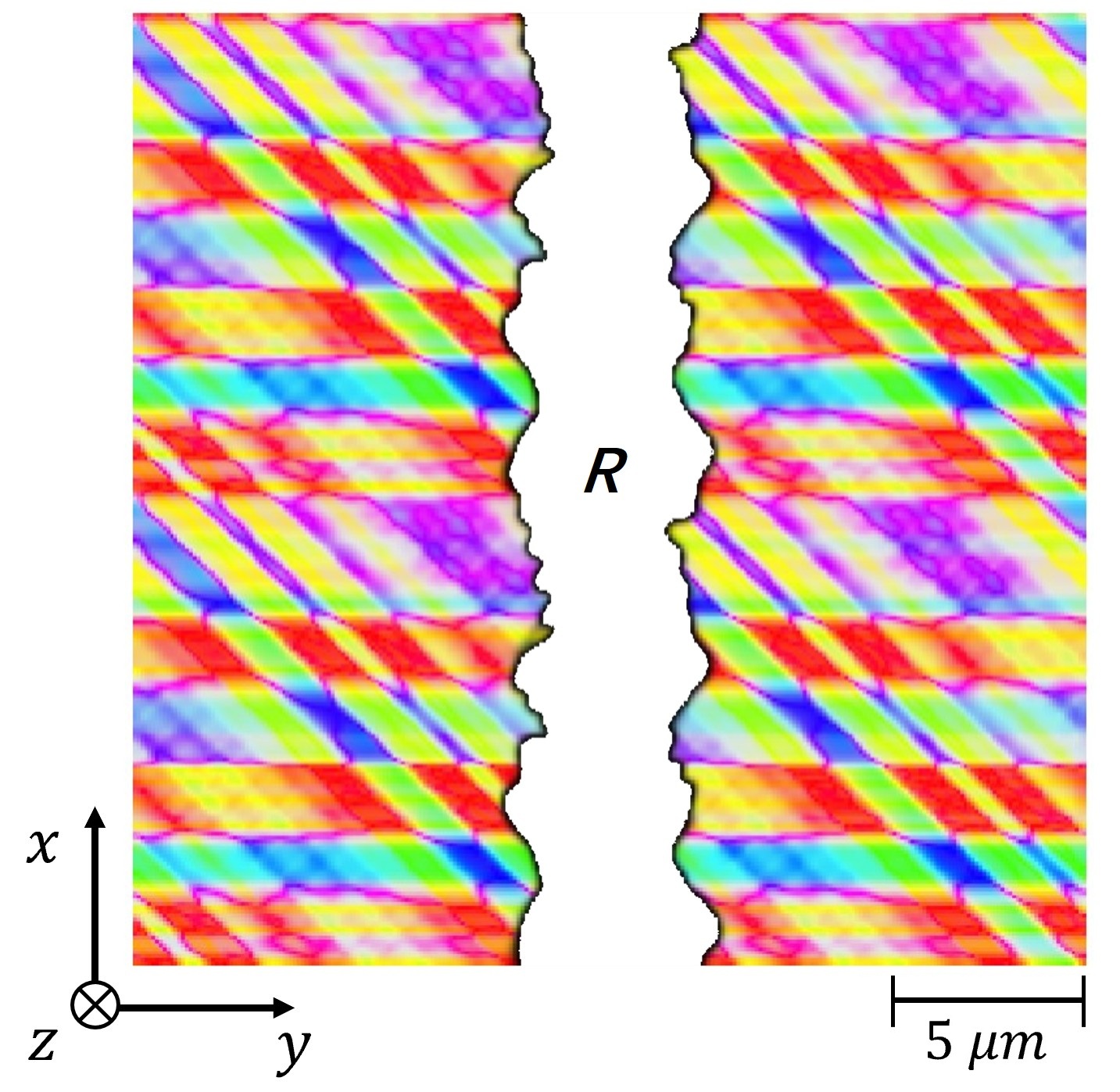}
    \hfill
    \includegraphics[width = .2\textwidth]{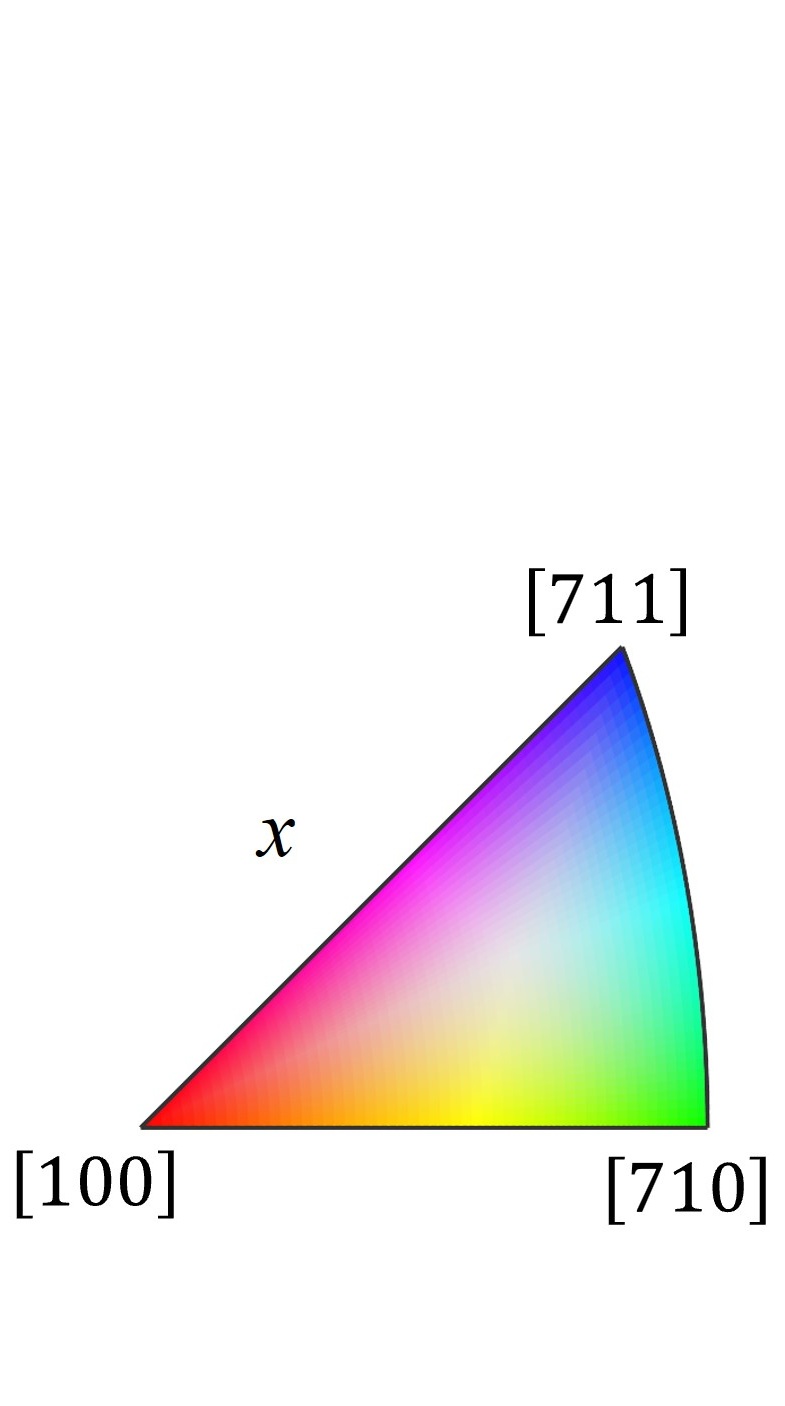}
    \hfill
    \includegraphics[width = .39\textwidth]{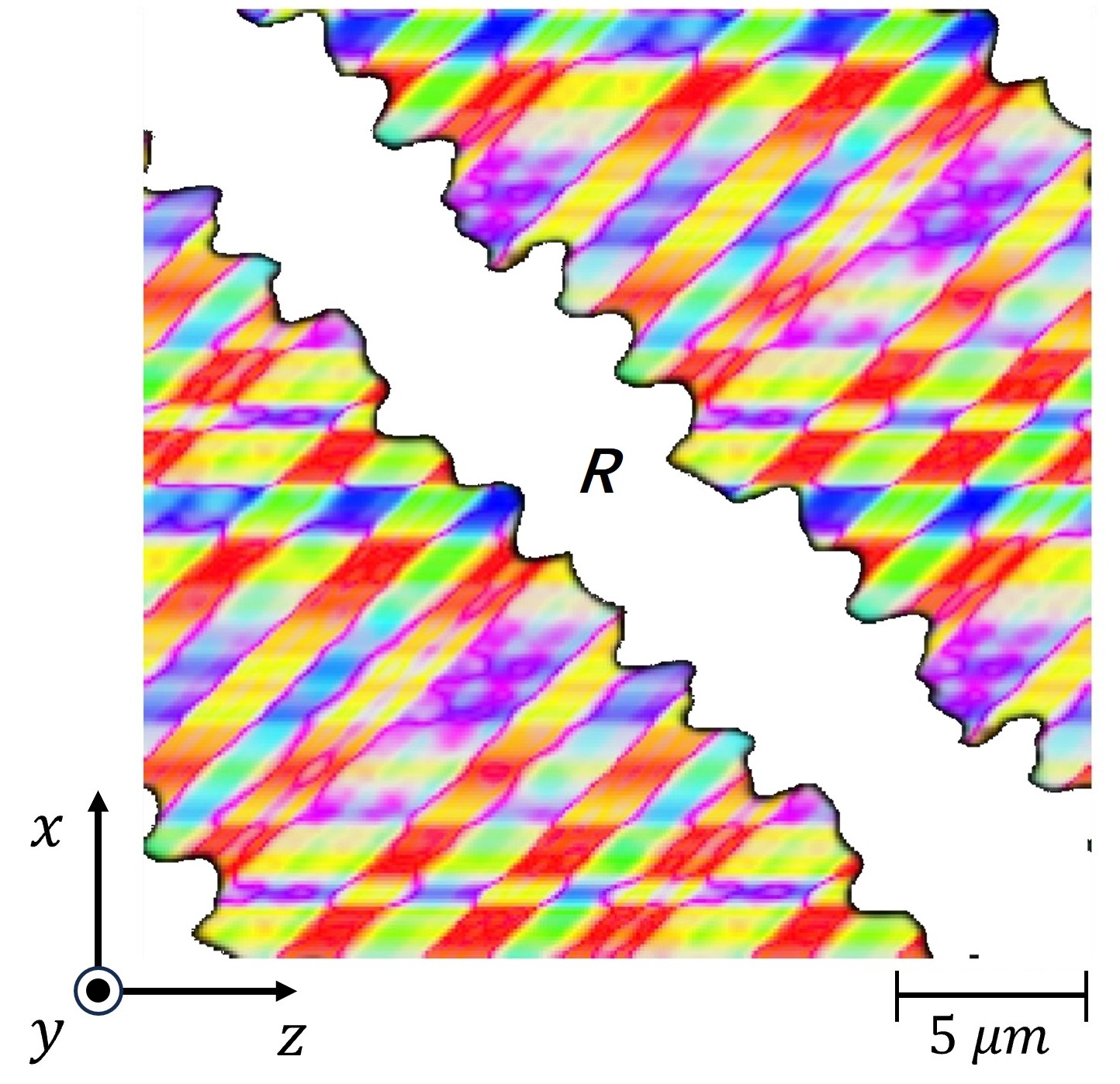}
    \caption{Intersection of recrystallization fronts with the [001] lattice plane; the recrystallized grain is shown in white; left: nucleus of the recrystallized grain oriented along the [010] plane, right: nucleus oriented along the [111] plane.}
    \label{fig:GBstraight}
\end{figure}
\begin{figure}[htb]  
    \centering\includegraphics[width = .45\textwidth]{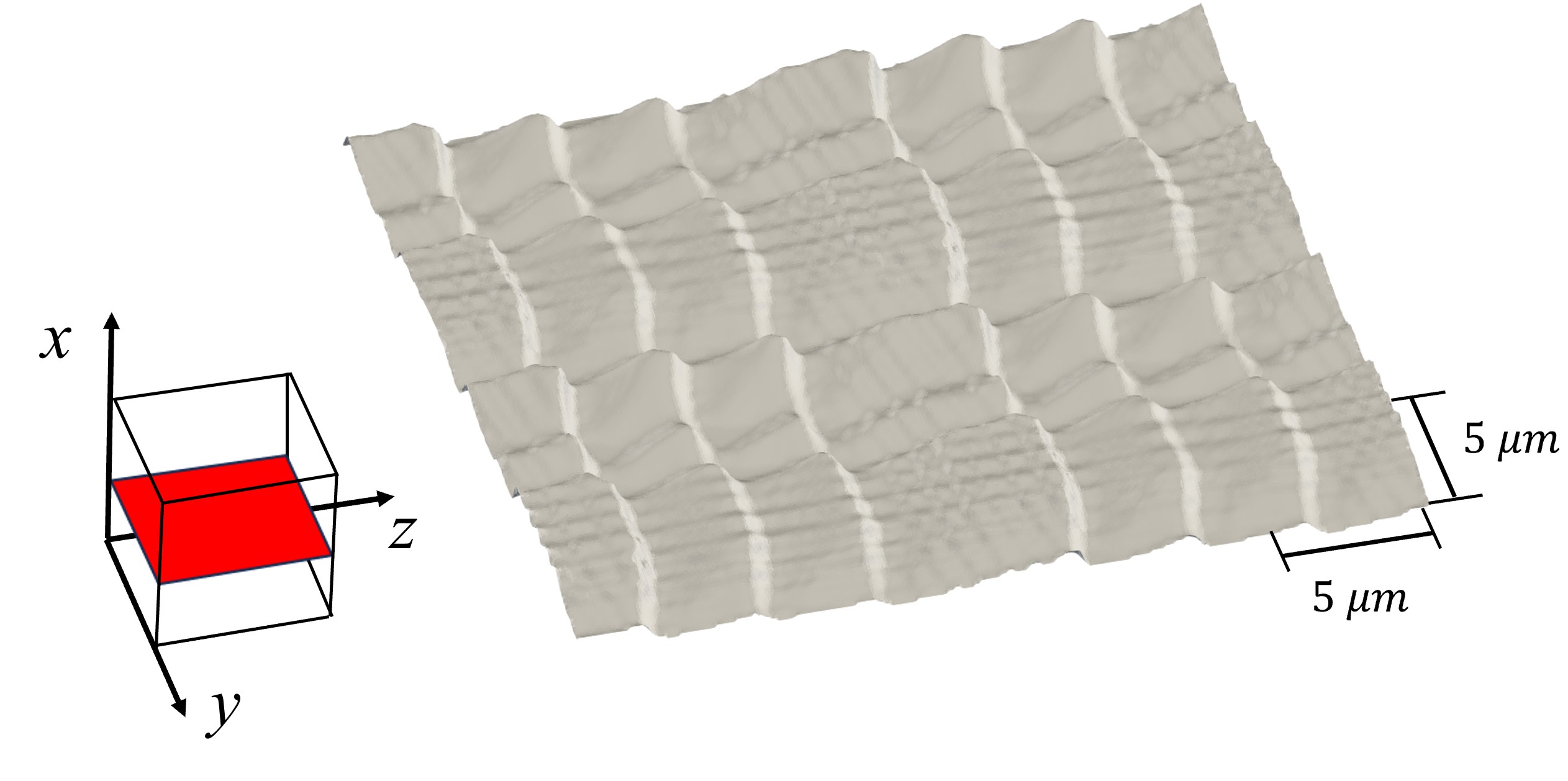}
    \hfill
    \includegraphics[width = .35\textwidth]{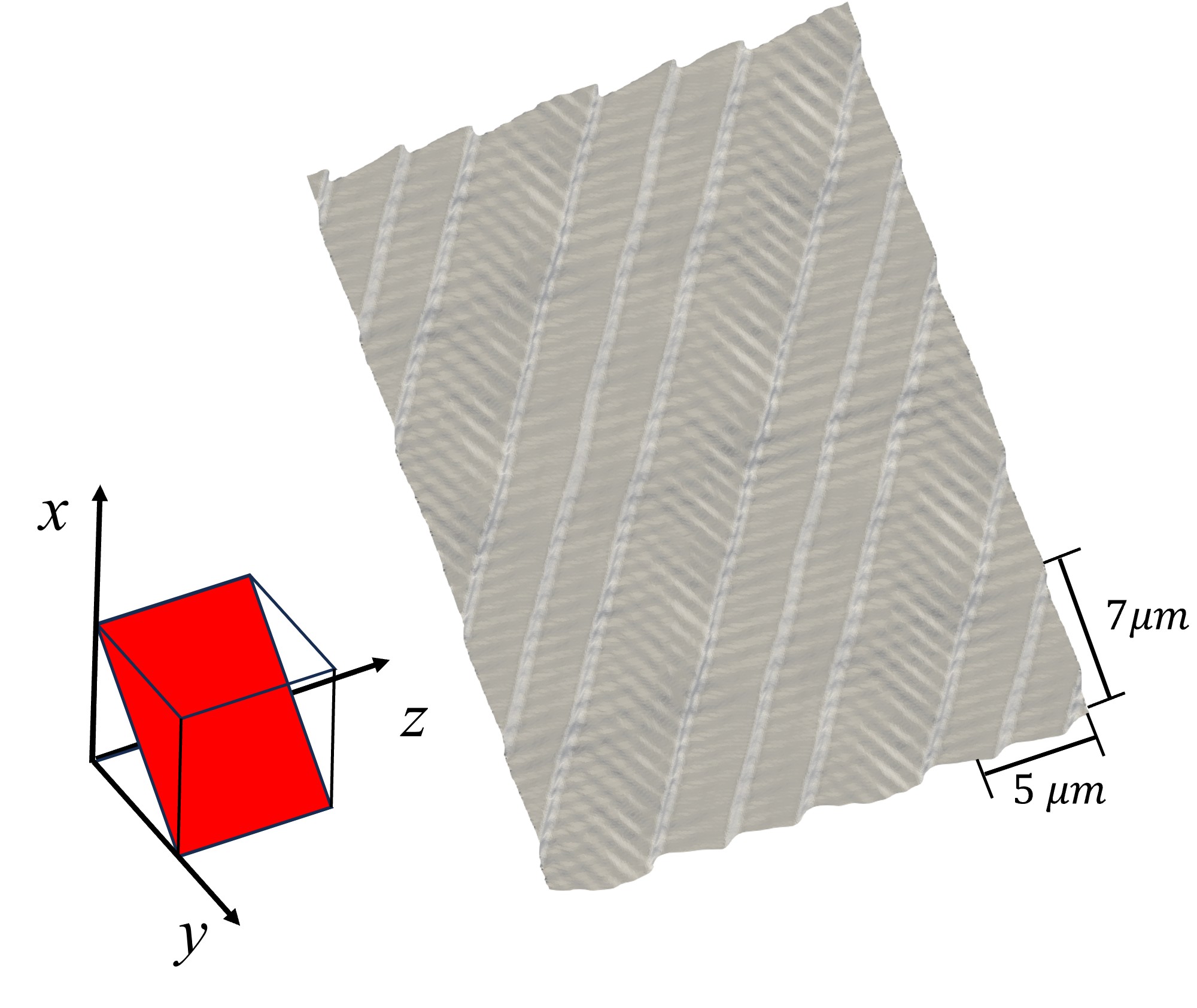}
    \\
    \includegraphics[width = .4\textwidth]{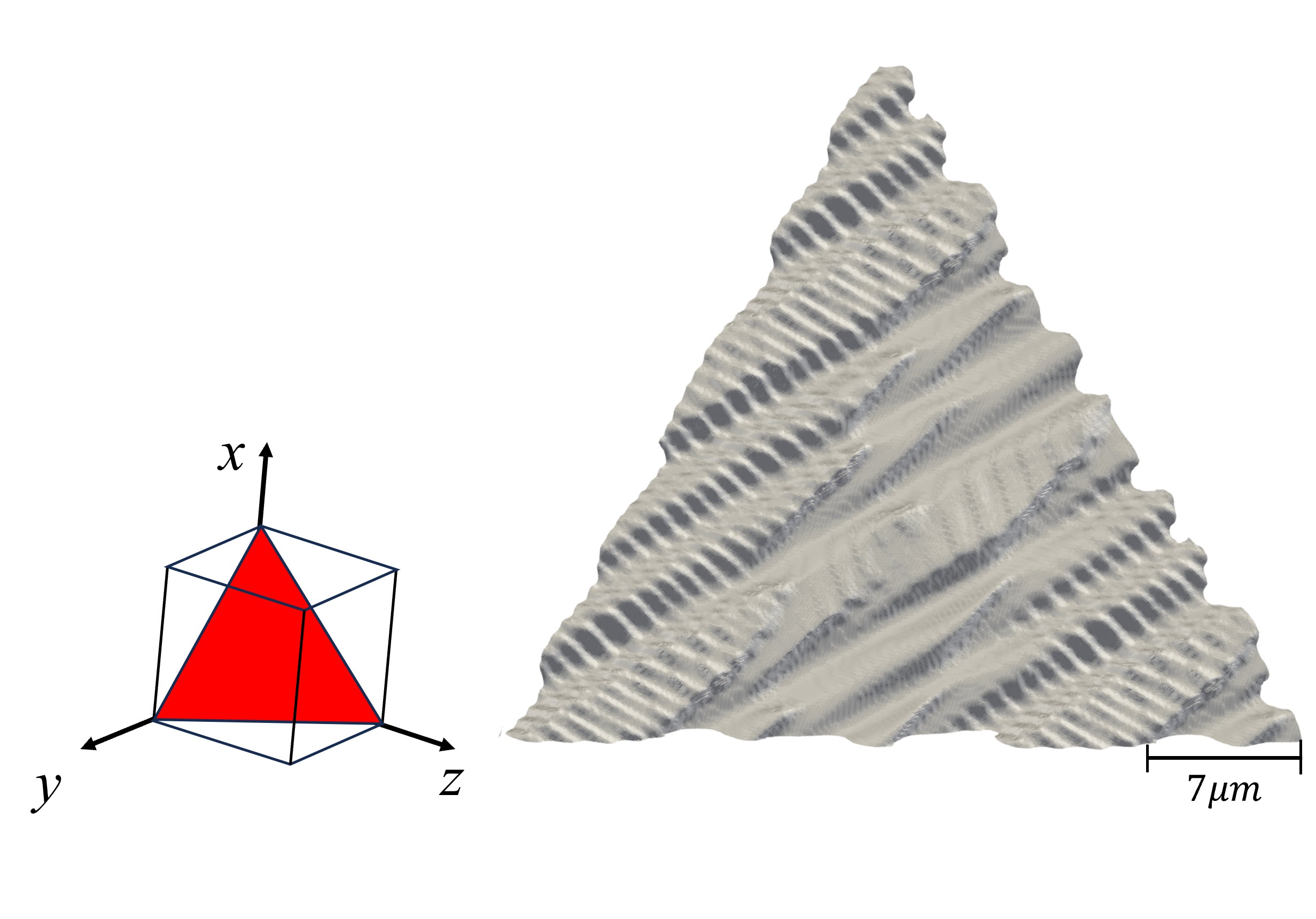}
    \hfill
    \includegraphics[width = .4\textwidth]{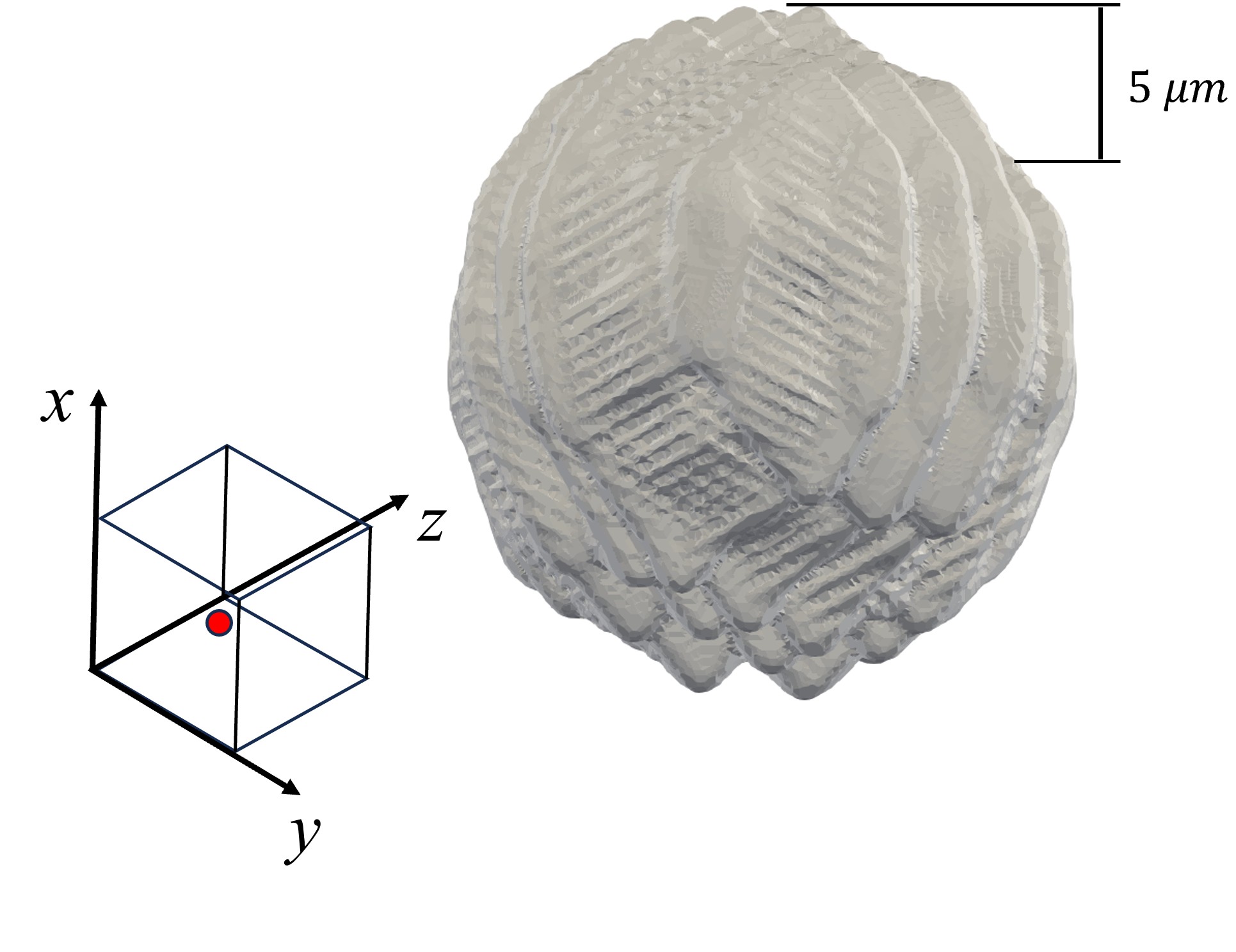}
    \caption{3D rendering of grain boundaries emerging during grain growth in the dislocation microstructure of \autoref{fig:cells4s}; the orientation of the respective grain nuclei is indicated by the red surfaces in the insets; top left: [100] oriented planar nucleus, top right: [110] oriented planar nucleus, bottom left: [111] oriented planar nucleus; bottom right: spherical nucleus.}
    \label{fig:GB3D}
\end{figure}
Results for [100] and [111] oriented nuclei are shown in \autoref{fig:GBstraight}. We see that the recrystallization front develops irregular oscillations whose morphology is essentially dictated by the underlying defect energy pattern and whose dominant wavelength matches the oscillations in crystal orientations, while much smaller and higher-frequency 'wigglings' reflect the presence of incidental dislocation boundaries (see \autoref{fig:GBstraight}, left). At the same time, it is clear that the anisotropic morphology of the underlying dislocation pattern has a discernable influence on grain boundary structure and grain growth. 
\begin{figure}[b]  \centering
    \includegraphics[width = .28\textwidth]{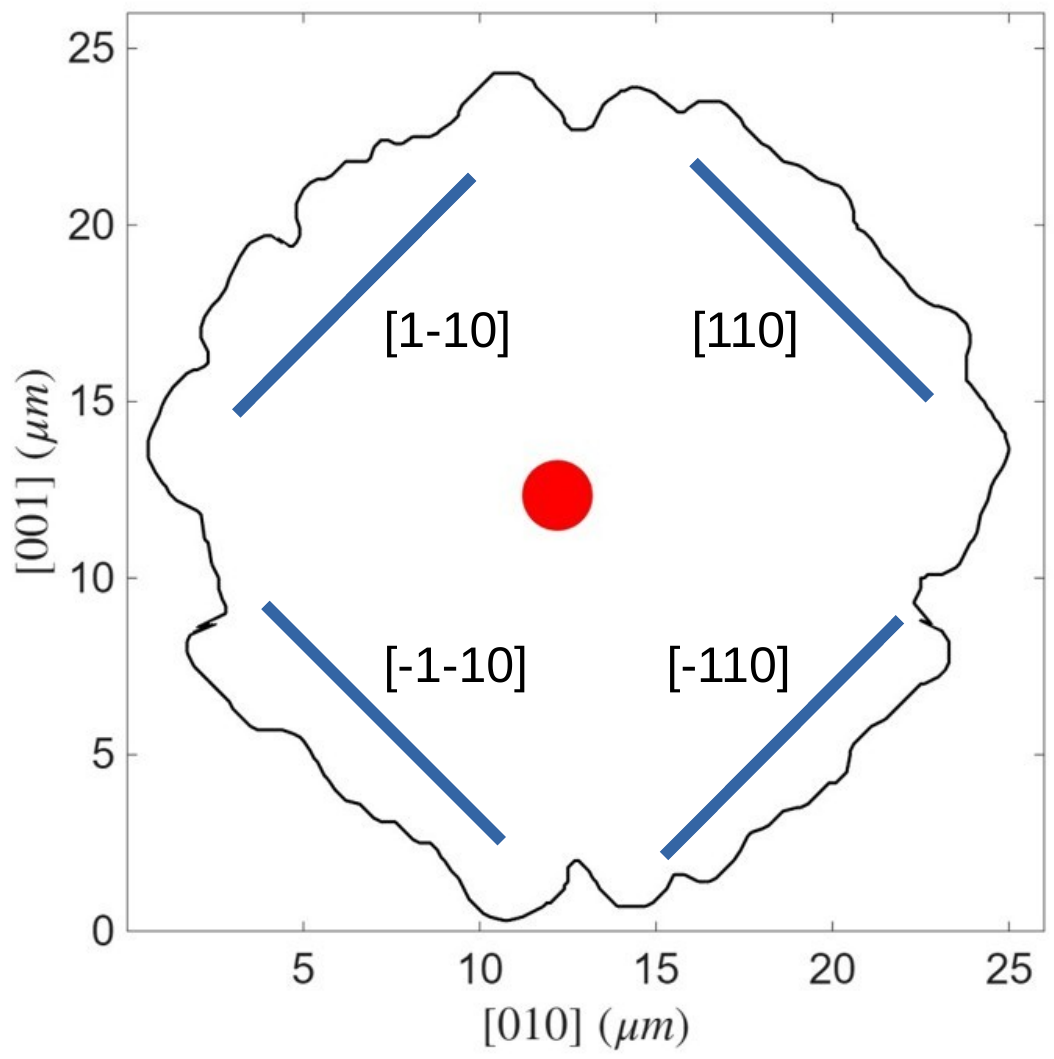}
    \includegraphics[width = .28\textwidth]{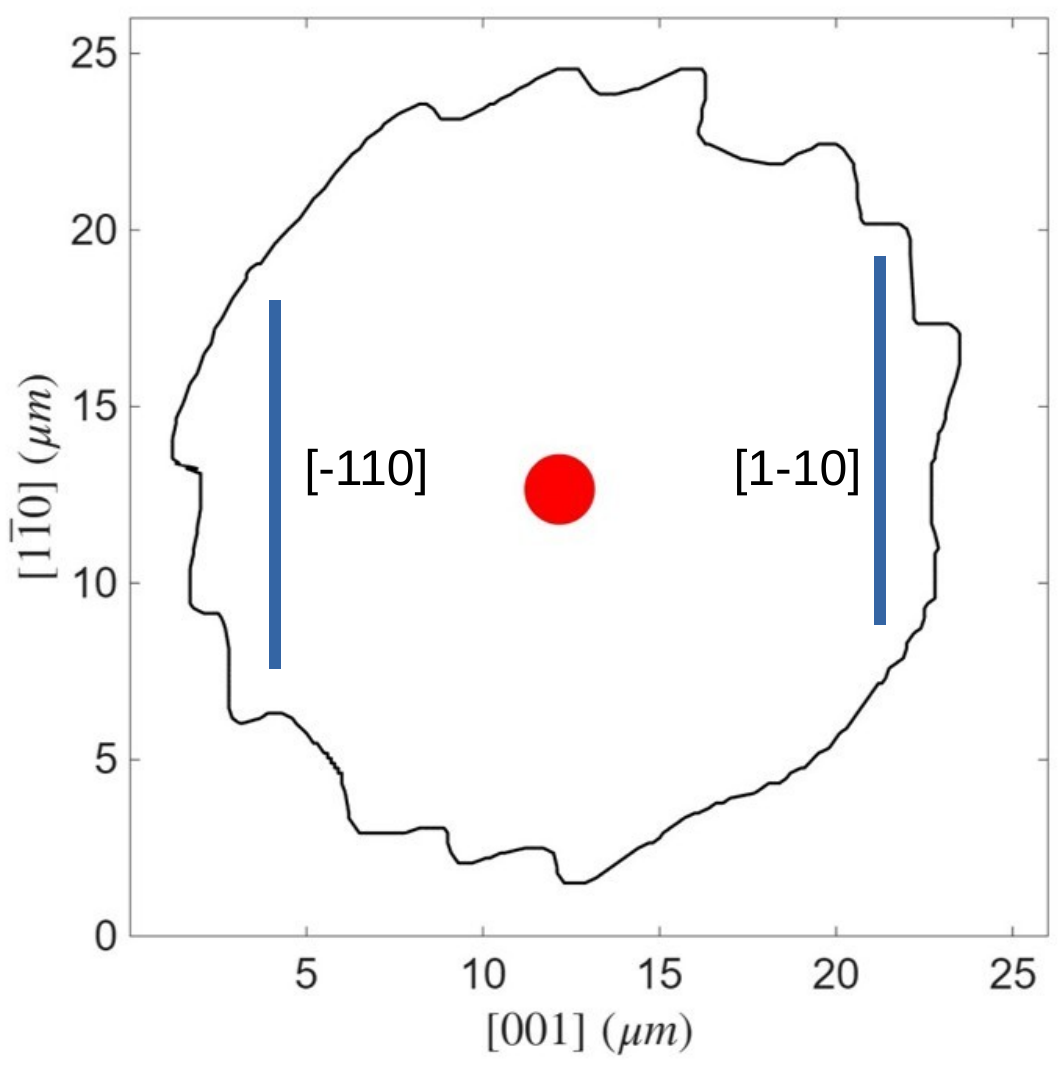}
    \includegraphics[width = .28\textwidth]{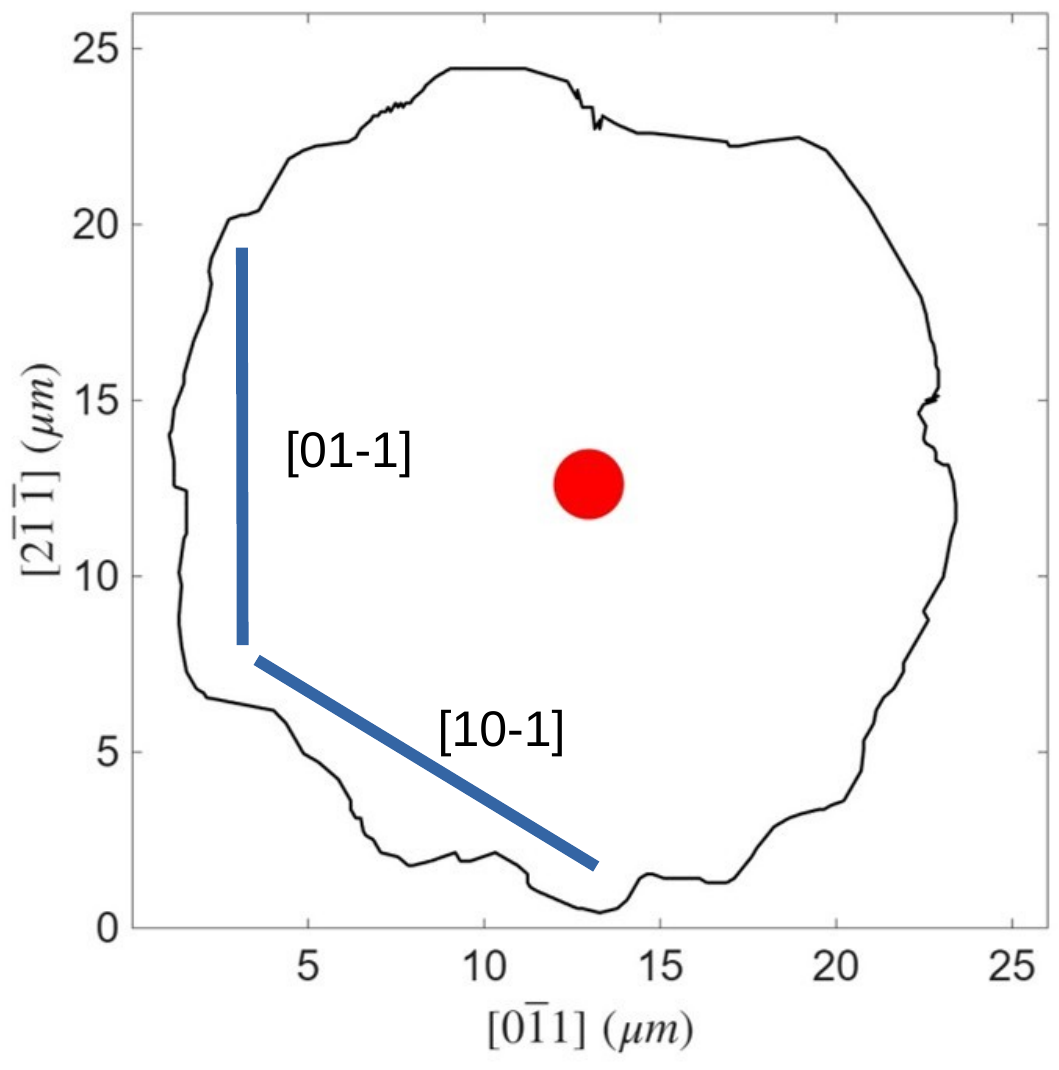}
    \caption{Central sections of recrystallized grain (3D image in  \autoref{fig:GB3D}, bottom right) grown from spherical nucleus (red circle), the GB trace in the section plane is shown in black; left: section in [100] plane, center: section in [110] plane, right: section in [111] plane;     the straight labeled lines indicate preferential faceting parallel to [110] planes.}
    \label{fig:GBnucleo}
\end{figure}

\begin{figure}[tb]  
    \includegraphics[width = .48\textwidth]{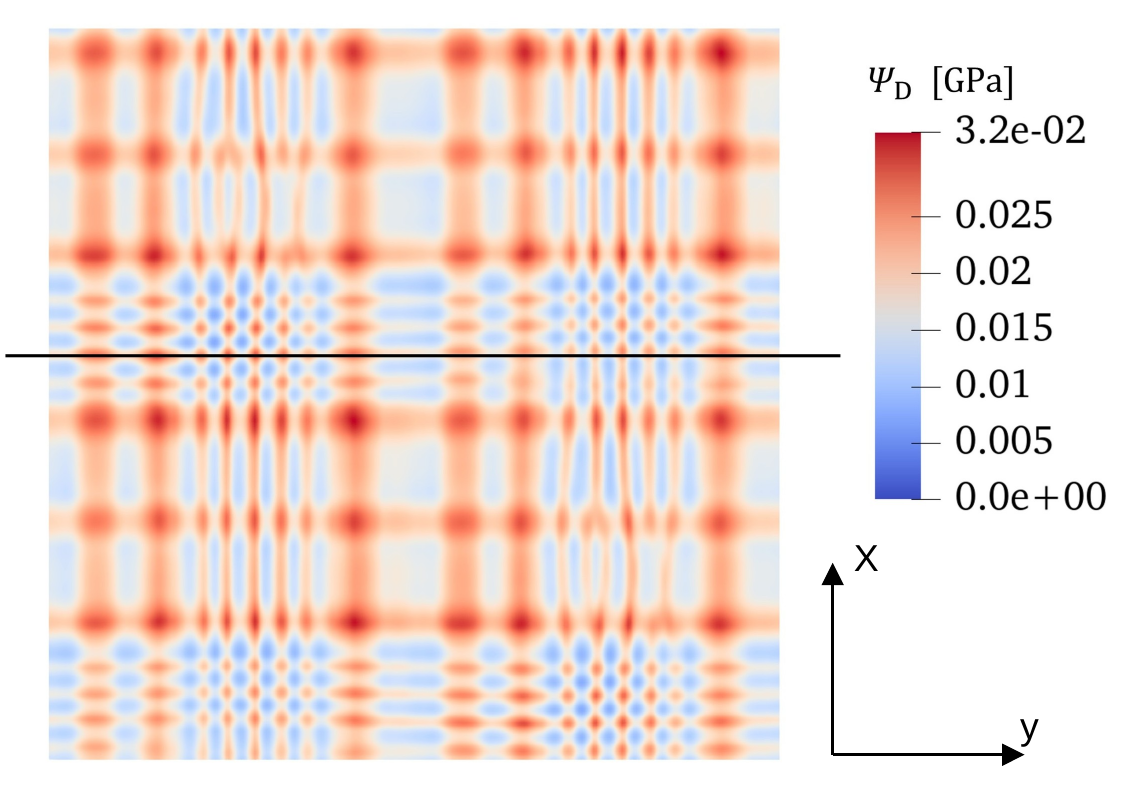}
    \hfill
    \includegraphics[width = .5\textwidth]{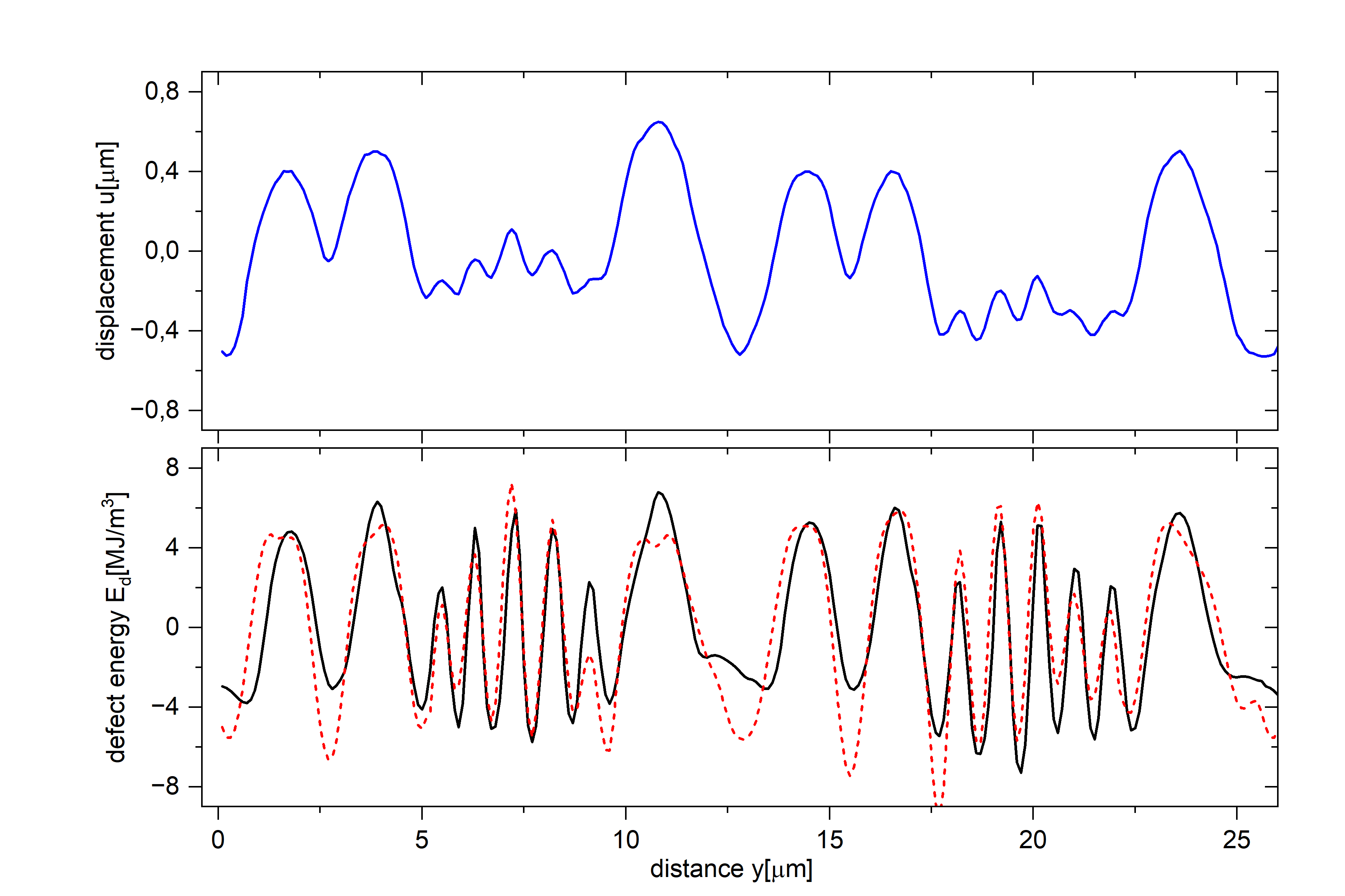}
    \caption{Left graph: defect energy pattern transversed by a [100] oriented GB; the walls are perpendicular to [100] and [010] directions, black line: average position of GB; right graph: GB displacement profile from black line (blue) and defect energy along black line (black); the red dashed line gives the defect energy evaluated from the displacement profile according to Eq. (36).}
    \label{fig:GBpattern}
\end{figure}

3D illustrations of the grain boundary morphology are given in \autoref{fig:GB3D} for GB orientations [010], [110], and [111], as well as for the  boundary of a recrystallized grain that emerged from a spherical nucleus that was initially placed in the center of the simulation supercell. Again we see characteristic variations in morphology depending on the orientation of the boundary relative to the anisotropic dislocation microstructure. The case of a spherical nucleus, which in a crystal with homogeneous defect energy would grow into a spherical shape of the recrystallized grain (note that we do not account for anisotropies in GB energy and/or mobility), vividly illustrates the efficiency of defect energy variations in promoting irregular grain shapes. The oscillations of the grain boundary exhibit a characteristic two-scale morphology which mirrors that of the dislocation pattern, with small short-wavelength undulations representing the effect of incidental boundaries and larger long-wavelength undulations associated with geometrically necessary boundaries as illustrated in \autoref{fig:GBpattern}. 

The velocity of the GB depends on its orientation relative to the inhomogeneous defect microstructure and is therefore anisotropic. For an initially spherical nucleus, this leads to an emergence of preferred GB orientations (faceting) as shown in \autoref{fig:GBnucleo} for the expanding grain illustrated in \autoref{fig:GB3D}, bottom right. In a cross section along the [100] plane, GB orientations near the [011] and [01-1] directions are preferred, indicating that these are the directions of slowest growth.


\section{Theoretical considerations}

To go beyond mere simulation and arrive at a deeper theoretical understanding of the morphology and anisotropic velocity of GBs in the present model, we move to a theoretical description where we switch from the PF framework to a description in terms of interface dynamics. We describe a macroscopically planar GB as a manifold $u(\Br)$ where $\Br = [y,z]$ is a vector in the plane of the initial GB nucleus and $u$ is the displacement of the GB in the direction perpendicular to this plane. In this framework, the evolution of the GB is described by the equation
\begin{equation}
    \frac{1}{M_{\rm g}} \partial_t u(\Br) = \sigma_{\rm g} \Delta u + \langle \Psi_{\rm D} \rangle + \delta \Psi_{\rm D}(\Br,u).
\end{equation}
This corresponds to the quenched Edwards-Wilkinson (EW) equation which has been extensively studied in the context of interface dynamics in disordered media. Here the GB dynamics is governed by the interplay between GB energy, which tries to flatten the GB, mean defect energy $\langle \Psi_{\rm D} \rangle$, which plays the role of a spatially homogeneous mean driving force, and defect energy fluctuations $\delta \Psi_{\rm D}(\Br,u)$, which tend to corrugate the GB and act as an effective pinning field that obstructs GB motion. Because these fluctuations in general depend on $u$, the quenched E-W equation is intrinsically nonlinear. 

We now focus first on the case of a [100] oriented GB. In this case, the energy 'landscape' traversed by the GB is illustrated in \autoref{fig:GBpattern}, left. The average position of the GB is indicated by a black line in \autoref{fig:GBpattern}, left, and the defect energy profile along this line together with the linear GB profile $u(y)$ are shown in \autoref{fig:GBpattern}, right. A conspicuous feature of the defect energy fluctuations experienced by a [100] oriented GB is that they can be approximated as a {\em multi-planar noise} of the form
\begin{equation}
\delta \Psi_{\rm D}(u,\Br) = \delta \psi_u (u) + \delta \psi_y(y) + \delta \psi_z(z) 
\end{equation}
where the different contributions represent independent random variables of zero mean. Each of the three contributions is strongly correlated in the directions parallel to one of the three [100] crystal lattice planes but only short range correlated in perpendicular direction. With this structure, we can approximately solve the E-W equation using the Ansatz
\begin{equation}
u(\Br) = \bar{u}(t) + u_y(y) + u_z(z)
\end{equation}
which leads to 
\begin{eqnarray}
\partial_t \bar{u} = 
    \frac{1}{M_{\rm g}} \partial_t \bar{u} = \langle \Psi_{\rm D} \rangle + \bar{\psi}_u(u)\;
    \label{eq:GBav},\\
    \sigma_{\rm g} \frac{\partial^2 u_y}{\partial y^2} = \psi_y(y)
    \;,\quad
    \sigma_{\rm g} \frac{\partial^2 u_z}{\partial z^2} = \psi_z(z)
    \;.
    \label{eq:GBprof}
\end{eqnarray}
In \autoref{eq:GBav}, the effective fluctuations $\bar{\psi}_u$ acting on the average GB position are obtained, for a given $\bar{u}$ by averaging $\psi_u$ over the fluctuations $u_y,u_z$ of the GB profile, that can be evaluated from \autoref{eq:GBprof} . 

According to \autoref{eq:GBprof}, the defect energy profiles in the $y$ and $z$ directions are related to the respective GB profiles by simple double differentiation. To demonstrate that this idea indeed provides an excellent description of the GB morphology, we have used it to calculate the energy profile along the black line in \autoref{fig:GBpattern} from the corresponding GB profile. The result is shown as red dashed line in \autoref{fig:GBpattern}, bottom right. The curve obtained from differentiation of the GB displacement field $u(\Br)$ is in excellent agreement with the actual energy profile, confirming that the GB can indeed be envisaged as an elastic manifold moving in a field of multi-planar disorder. Moreover, the idea that the shape of the moving GB is controlled by energy fluctuations that are strongly correlated along the direction of motion of the GB is consistent with experimental observations, see \autoref{fig:recrystallization}.

\begin{figure}[tb] 
\centering
    \includegraphics[width = .95\textwidth]{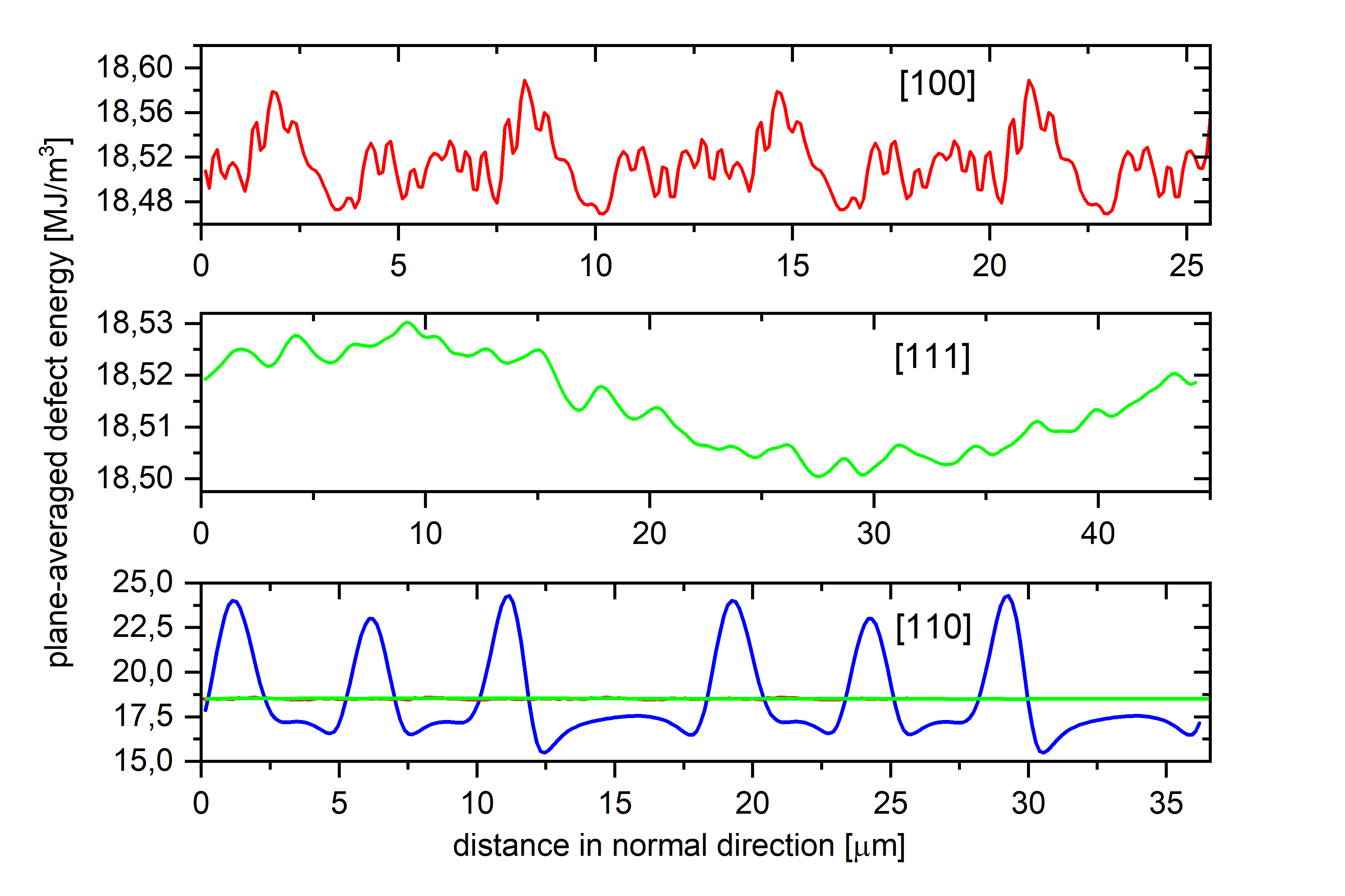}
    \caption{Planar averages of the defect energy, as functions of the coordinate perpendicular to the averaging plane; top: [100] plane, center: [111] plane, bottom: [110] plane; note the different scales; the bottom graph contains the averaged energies for [100] and [111] orientations (red and green) for comparison.}
    \label{fig:GBenergies}
\end{figure}
To understand the differences in GB mobility that give rise to faceting of an initially spherical GB nucleus, we look at the transversally averaged defect energy $\langle \delta \Phi_{\rm D}(u)\rangle_{\Br}$ where the angular brackets $\langle \dots \rangle_{\Br}$ denote averages over the directions parallel to the mean GB plane. Results are shown in 
\autoref{fig:GBenergies}. These curves demonstrate that the average fluctuations of the transversally averaged defect energy are very low for GBs with [100] and [111] orientations. However, appreciable fluctuations are experienced by GBs with [110] orientation. This can be understood by observing that the 'geometrically necessary' dislocation walls are preferentially aligned with [110] planes, thus, [100] oriented GBs experience strong fluctuations of their average energy which may reduce their effective mobility. GBs of oblique orientation, on the other hand, average over these fluctuations and thus experience reduced pinning. We note that similar observations were reported by \citet{yadav2021influence} in the context of a 2D model with artificially introduced energy fluctuations. 

\section{Discussion and conclusions}

We have, for the first time, developed a model that can without ad-hoc assumptions describe the effect of heterogeneous dislocation microstructure on grain boundary motion during recrystallization. Unlike previous studies using a similar recrystallization model, the present work derives the defect energy from a model of the formation of heterogeneous dislocation microstructures, that captures the emergence of incidental and geometrically necessary dislocation booundaries. We also note that, as opposed to prevoius works which consider a 2D framework, the present formulation is fully three-dimensional. 

While the present model illustrates the generic framework for the case of a recrystallized grain expanding into a deformed single crystal, it is (barring limitations of computational cost) straightforward to generalize this framework to polycrystals where the microstructure morphology changes from grain to grain. An important restriction of the present formulation is, however, that the simulations envisage situations where either the grain microstructure or the dislocation microstructure are static: Dislocations evolve in a static grain microstructure (specifically a single grain) and grain boundaries then move at fixed dislocation microstructure such that any changes in the dislocation densities arise only as a consequence of the crystallographic changes associated with the re-structuring of the crystal lattice. 

The difficulties associated with maintaining a correct balance of dislocation density changes associated with dislocation fluxes, and dislocation microstructure changes associated with crystallographic change, need to be addressed in order to be able to apply the present framework to the important problem of dynamic recrystallization. We leave such a study for future investigation.

\section*{Acknowledgement}
We gratefully acknowledge joint support by DFG and CSC under grant no. M-5011, and support by DFG under grant no Za 171/13-1.

\bibliographystyle{spbasic}       
\bibliography{phasefield.bib}

\begin{appendix}
\section*{Appendix: Equivalence of dislocation density degradation and energy degradation}

We start with \autoref{eq:disdegrad} which describe the degradation of the slip system dislocation densities. We consider the case of a monotonically decreasing volume fraction $f$ of the parent grain:
\begin{equation}
    \frac{\partial_t \rho}{\rho}=\frac{\partial_t \rho^{\beta}}{\rho^{\beta}} = 
    \frac{\partial_t f}{f} \quad,\quad 
    \partial_t (\Brho^{\beta}.\Brho^{\beta})
 = \frac{\partial_t \rho}{\rho}\left(\Brho^{\beta}.\Brho^{\beta}  - \frac{2A}{nD} \rho^2 \right). 
 \label{eq:dtrho}
\end{equation}
and use the definition of the defect energy, 
\begin{equation}
\Psi_{\rm D} = \sum_{\beta}\left(A G b^2 \rho^{\beta} \ln \left(\rho {\ell}^2\right) + D G b^2 \frac{\Brho^{\beta}\cdot \Brho^{\beta}}{2 \rho}\right) = A G b^2 \rho \ln\left(\rho{\ell}^2\right)+\sum_{\beta} D G b^2 \frac{\Brho^{\beta}\cdot \Brho^{\beta}}{2 \rho} .
\label{eq:energycdd2a}
\end{equation}
In absence of dislocation density changes related to dislocation motion, the evolution of the defect energy is then given by 
\begin{eqnarray}
\frac{\partial\Psi_{\rm D}}{\partial t} &=& A G b^2 \rho \left(\ln\left(\rho {\ell}^2 \right) +1\right)\frac{\partial_t \rho}{\rho}  - D G b^2 \left[\sum_{\beta} \frac{\Brho^{\beta}\cdot \Brho^{\beta}}{2 \rho}\right]\frac{\partial_t \rho}{\rho}\nonumber\\
&+& D G b^2 \sum_{\beta} \left(\frac{\Brho^{\beta}\cdot \Brho^{\beta}}{ \rho}-\frac{A\rho}{nD} \right)\frac{\partial_t \rho}{\rho}  .
\label{eq:energycdd2b}
\end{eqnarray}
We see that this simplifies to
\begin{equation}
\frac{\partial\Psi_{\rm D}}{\partial t} = \left[A G b^2 \rho \ln\left(\rho{\ell}^2\right)+ D G b^2 \sum_{\beta} \frac{\Brho^{\beta}\cdot \Brho^{\beta}}{ 2\rho}\right]\frac{\partial_t \rho}{\rho}  = \Psi_{\rm D} \frac{\partial_t \rho}{\rho}.
\label{eq:energycdd2c}
\end{equation}
Using \autoref{eq:dtrho}, we see that this can be interpreted as a differential equation 
\begin{equation}
\frac{\partial_t \Psi_{\rm D}}{\Psi_{\rm D}} = \frac{\partial_t f}{f}\quad\rightarrow \quad
\frac{\partial \Psi_{\rm D}}{\partial f} =\frac{\Psi_{\rm D}}{f}  
\label{eq:dfrho}
\end{equation}
which with the initial conditions $\Psi_{\rm D} = \Psi_{\rm D}^0$, $f=1$ integrates to 
\begin{equation}
\Psi_{\rm D} = \Psi_{\rm D}^0 f.
\end{equation}
Thus, $f$ acts as a degradation function for the defect energy present prior to recrystallization.
\end{appendix}

\end{document}